\documentclass{elsart}
\usepackage{epsfig}
\usepackage{amssymb}
\usepackage{amsmath}
\usepackage{bm}
\usepackage{graphicx}
\usepackage{graphics}
\usepackage{color}
\usepackage{pstricks}
\usepackage{cite}
\usepackage{lscape}
\usepackage[leftbars]{changebar}
\newcommand{\bomega}{\bm \omega}

\newcommand{\fuf}{\underline{\overline{\mathbf u}}}

\newcommand{\fpf}{\underline{\overline{p}}}

\newcommand{\dd}{\text{d}}
\newcommand{\bM}{{\mathbf M}}
\newcommand{\bC}{{\mathbf C}}

\newcommand{\bD}{{\mathbf D}}

\newcommand{\bA}{{\mathbf A}}
\newcommand{\bS}{{\mathbf S}}

\newcommand{\Cd}{C_{\dd}}

\newcommand{\cP}{{\mathcal P}}
\newcommand{\cI}{{\mathcal I}}
\newcommand{\cF}{{\mathcal F}}
\newcommand{\cQ}{{\mathcal Q}}
\newcommand{\cT}{{\mathcal T}}
\newcommand{\cL}{{\mathcal L}}

\newcommand{\cH}{{\mathcal H}}
\newcommand{\cG}{{\mathcal G}}
\newcommand{\cK}{{\mathcal K}}

\newcommand {\bnabla}      {\bm\nabla}
\newcommand {\btau}        {\bm\tau}
\newcommand {\ou}          {\overline{u}}
\newcommand {\op}          {\overline{p}}
\newcommand {\tou}         {\widetilde{\overline{u}}}

\newcommand {\uia}         {u_{i}^{\ast}}
\newcommand {\ouia}        {\overline{u_{i}^{\ast}}}

\newcommand {\touia}       {\widetilde{\overline{u_{i}^{\ast}}}}
\newcommand {\uja}         {u_{j}^{\ast}}
\newcommand {\ouja}        {\overline{u_{j}^{\ast}}}
\newcommand {\touja}       {\widetilde{\overline{u_{j}^{\ast}}}}
\newcommand {\bu}          {\mathbf{u}}
\newcommand {\bs}          {\mathbf{s}}
\newcommand {\obu}         {\overline{\mathbf{u}}}
\newcommand {\tobu}        {\widetilde{\overline{\mathbf{u}}}}
\newcommand {\bua}         {\mathbf{u^{\ast}}}

\makeatletter
\DeclareRobustCommand{\Cpp}
{\valign{\vfil\hbox{##}\vfil\cr
   \textsf{C\kern-.1em}\cr
   $\hbox{\fontsize{\sf@size}{0}\textbf{+\kern-0.05em+}}$\cr}%
}
\makeatother

\makeatletter
\DeclareRobustCommand{\Cp}
{\valign{\vfil\hbox{##}\vfil\cr
   \textsf{C\kern-.1em}\cr
   $\hbox{\fontsize{\sf@size}{0}\textbf{\kern-0.05em}}$\cr}%
}
\makeatother
\journal{\it J. Comput. Phys.}
\volume{224}
\issue{1}
\pubyear{2007}
\firstpage{241}
\lastpage{266}

\begin{document}

\begin{frontmatter}
\title{A coupled approximate deconvolution\\ and dynamic mixed scale model\\ for large-eddy simulation}
\author[EPFL]{Marc A. Habisreutinger},
\ead{marc-antoine.habisreutinger@epfl.ch}
\author[EPFL]{Roland Bouffanais\corauthref{cor}\thanksref{FNRS}},
\corauth[cor]{Corresponding author.}
\thanks[FNRS]{Supported by a Swiss National Science Foundation Grant No. 200020--101707}
\ead{roland.bouffanais@epfl.ch}
\author[EPFL]{Emmanuel Leriche},
\ead{emmanuel.leriche@epfl.ch}
\author[EPFL]{Michel O. Deville}
\ead{michel.deville@epfl.ch}
\address[EPFL]{Laboratory of Computational Engineering,\\ \'Ecole Polytechnique F\'ed\'erale de Lausanne,\\ STI -- ISE -- LIN, Station 9,\\ CH--1015 Lausanne, Switzerland}
\begin{abstract}
Large-eddy simulations of incompressible Newtonian fluid flows with approximate deconvolution models based on the van Cittert method are reported. The Legendre spectral element method is used for the spatial discretization to solve the filtered Navier--Stokes equations. A novel variant of approximate deconvolution models blended with a mixed scale model using a dynamic evaluation of the subgrid-viscosity constant is proposed. This model is validated by comparing the large-eddy simulation with the direct numerical simulation of the flow in a lid-driven cubical cavity, performed at a Reynolds number of 12'000. Subgrid modeling in the case of a flow with coexisting laminar, transitional and turbulent zones such as the lid-driven cubical cavity flow represents a challenging problem. Moreover, the coupling with the spectral element method having very low numerical dissipation and dispersion builds a well suited framework to analyze the efficiency of a subgrid model. First- and second-order statistics obtained using this new model are showing very good agreement with the direct numerical simulation. Filtering operations rely on an invertible filter applied in a modal basis and preserving the $C^0$-continuity across elements. No clipping on dynamic parameters was needed to preserve numerical stability.
\end{abstract}
\end{frontmatter}
\section{Introduction}
Large-eddy simulation (LES) represents a way of reducing the number of degrees of freedom of the simulation with respect to the requirements of the direct numerical simulation (DNS). This is done by calculating only low-frequency modes in space and modeling high-frequency ones, the scale separation being performed by filtering in space the Navier--Stokes equations. Large-scale structures are obtained by the computed flow dynamics while the behavior of subgrid scales and their interaction with large eddies are modeled by the additional term in the LES governing equations resulting from filtering the Navier--Stokes equations. The expression of the additional term as a function of the resolved field is referred to as subgrid modeling.

Approximate deconvolution models (ADM) constitute a particular family of subgrid models. They rely on the attempt to recover, at least partially, the original unfiltered fields, up to the grid level, by inverting the filtering operator applied to the Navier--Stokes equations. The focus here is on the approximate iterative method introduced by Stolz and Adams \cite{stoltz99} which is based on the van Cittert procedure. This method was subsequently applied to incompressible wall-bounded flows \cite{stolz05:_high}, to compressible flows and to shock-boundary layer interaction \cite{stolz01:compressible} using a new variant ADM-RT, blending ADM with a relaxation term (RT) increasing the dissipative character of the model. Transitional flows were also investigated by Schlatter \etal\ \cite{schlatter04:_les}. Over the past five years, ADM spread over various fields of application. Gullbrand and Chow studied the effect of explicit filtering in the case of channel flow \cite{gullbrand03}. ADM were also more recently applied to the LES of a rectangular jet and to computational aero-acoustics by Rembold and Kleiser \cite{rembold04:_noise}. Particle-laden turbulent flows were investigated in the ADM framework by Shotorban and Mashayek \cite{shotorban05:_model}. From the numerical viewpoint, Schlatter \etal\ \cite{schlatter04:_les} used a parallel implementation of a mixed Fourier-Chebyshev spectral method. These models were also implemented in a finite volume framework in the semi-industrial code NSMB, Navier--Stokes Multi--Block, by von Kaenel \etal\ who applied it to shock-boundary layer interaction and channel flow in \cite{kaenel03:_large,kaenel04:_large}. To our knowledge, the only implementation based on the spectral element method (SEM) is due to Iliescu and Fischer \cite{iliescu03:_large} who used ADM based on the rational LES model (RLES) instead of the van Cittert one. More recently, Pruett \etal\ proposed a temporal ADM for LES \cite{pruett06} and a stability analysis of the LES-ADM equations was performed by Dunca and Epshteyn \cite{dunca06:_stolz_adams}.

LES of Newtonian incompressible fluid flows with ADM based on the van Cittert method using Legendre-SEM\ as spatial discretization to solve the filtered Navier--Stokes equations are envisaged for the first time in this paper. Following the idea of Winckelmans \etal\ \cite{winckelmans01:_asses_les} who coupled the ADM based on the van Cittert method and the Smagorinsky model \cite{smagorinsky63:_gener}, and Gullbrand and Chow \cite{gullbrand03} who proposed a dynamic version of the previous model, a new variant which blends ADM and the mixed scale model introduced by Sagaut \cite{sagaut96:_numer} with a dynamic evaluation of the subgrid-viscosity constant based on a Germano--Lilly type of procedure \cite{germano91,lilly92:_german} is proposed.

A specific filtering operation adapted to SEM\ and preserving continuity across elements is applied in a modal basis which was proposed in the $p$-version of finite elements and first used by Boyd \cite{boyd98:_two_cheby_legen} as a filtering technique. Depending on the transfer function, this filter is not projective and is therefore invertible, this property being essential for the deconvolution procedure. 

A DNS of the flow in a lid-driven cubical cavity performed at Reynolds number of 12'000 with a Chebyshev collocation method due to Leriche and Gavrilakis \cite{leriche00:_direc} is taken as the reference solution to validate the new model. Subgrid modeling in the case of a flow with coexisting laminar, transitional and turbulent zones such as the lid-driven cubical cavity flow represents a challenging problem. As the flow is confined and recirculating, any under- or over-dissipative character of the subgrid model can be clearly identified. Moreover, the very low dissipation and dispersion induced by SEM\ allows a pertinent analysis of the energetic action induced by any subgrid model, which is not feasible in the framework of low-order numerical methods. The coupling of the lid-driven cubical cavity flow problem with the SEM\ builds therefore a well suited framework to analyze the accuracy of the newly defined subgrid model. Bouffanais \etal\ in \cite{bouffanais05:_large,bouffanais06:_large} have performed LES of the flow in a lid-driven cubical cavity at a Reynolds number of 12'000 using the same physical parameters as the DNS from Leriche and Gavrilakis \cite{leriche00:_direc}. The numerical framework of \cite{bouffanais05:_large,bouffanais06:_large} is the same as the one used in the present article. Standard subgrid models were used in \cite{bouffanais05:_large,bouffanais06:_large}: dynamic Smagorinsky \cite{lilly92:_german,germano91} or dynamic mixed models \cite{zang93}.

The paper is organized as follows. In Section 2, the filtered Navier--Stokes equations are given, followed by a brief description of the space-time discretization using the spectral element method. The subgrid modeling is dealt in details in Section 3 and the numerical filters are described in Section 4. LES of the flow in the lid-driven cubical cavity, based on the subgrid models introduced in Section 3, is presented and thoroughly analyzed in Section 5. Finally, in Section 6 we present the conclusions.

\section{Governing equations and numerical method}
\subsection{Governing equations}
In the case of isothermal flows of Newtonian incompressible fluids, the LES governing equations for the filtered quantities denoted by an \textit{overbar}, obtained by applying a convolution filter $\cG \star$ to the Navier--Stokes equations, read
\begin{align}
&\frac{\partial\ou_{i}}{\partial t}+\frac{\partial}{\partial
  x_{j}}(\ou_{i}\ou_{j})=-\frac{\partial\overline{p}}{\partial
  x_{i}}+\nu\frac{\partial}{\partial x_{j}}\left(\frac{\partial\ou_{i}}{\partial x_{j}}+\frac{\partial\overline{u}_{j}}{\partial x_{i}}\right)-\frac{\partial\tau_{ij}}{\partial x_{j}},\label{eq:02-04}\\
&\frac{\partial \ou_j}{\partial x_j}  = 0,\label{eq:02-04bis}
\end{align}
the filtered velocity field $\obu=\cG \star \bu$ satisfying the divergence-free condition \eqref{eq:02-04bis} through the filtered reduced pressure field $\op$. The components of the subgrid tensor $\btau$ are given by
\begin{equation}\label{eq:deftau}
 \tau_{ij}=\overline{u_{i}u}_{j}-\ou_{i}\ou_{j},
\end{equation}
and $\nu$ is the kinematic viscosity. The closure of the filtered momentum equation \eqref{eq:02-04} requires $\btau$ to be expressed in terms of the filtered field which reflects the subgrid scales modeling and the interaction among all space scales of the solution.

\subsection{Space discretization}
The numerical method treats Eqs. \eqref{eq:02-04}--\eqref{eq:02-04bis} within the weak Galerkin formulation framework. The SEM\ consists in dividing the computational domain into a given number of spectral elements. In each spectral element, the velocity and pressure fields are approximated using Lagrange--Legendre polynomial interpolants. The reader is referred to the monograph by Deville \etal\ \cite{deville02:_high} for full details. The velocity and pressure are expressed in the $\mathbb{P}_p-\mathbb{P}_{p-2}$ functional spaces where $\mathbb{P}_p$ is the set of polynomials of degree lower than $p$ in each space direction. This spectral element method avoids the presence of spurious pressure modes as it was proved by Maday and Patera \cite{maday92:_nimes_n_stokes,maday89:_spect_navier_stokes}. The quadrature rules are based on a Gauss--Lobatto--Legendre (GLL) grid for the velocity nodes and a Gauss--Legendre grid (GL) for the pressure nodes. 

Borrowing the notation from Deville \etal\ \cite{deville02:_high}, the semi-discrete filtered Navier--Stokes equations resulting from space discretization are 
\begin{align}
\bM \frac{\dd \fuf}{\dd t}+ \bC \fuf +\nu \bA \fuf -\bD^T \fpf+\bD \underline{\btau}&=0,\label{eq:odes}\\
-\bD\fuf &=0\label{eq:constr}.
\end{align}
The diagonal mass matrix $\bM$ is composed of three blocks, namely the mass matrices $M$. The global vector $\fuf$ contains all the nodal velocity components while $\fpf$ is made of all nodal pressures. The matrices $\bA$, $\bD^T$, $\bD$ are the discrete Laplacian, gradient and divergence operators, respectively. The matrix operator $\bC$ represents the action of the nonlinear term written in convective form $\fuf\cdot \bnabla$, on the velocity field and depends on $\fuf$ itself. The semi-discrete equations constitute a set of nonlinear ordinary differential equations \eqref{eq:odes} subject to the weak incompressibility condition \eqref{eq:constr}.
\subsection{Time integration}\label{sec:time-integration}
Standard time integrators in the SEM\ framework handle the viscous linear term and the pressure implicitly by a backward differentiation formula of order $2$ (BDF2) to avoid stability restrictions such that $ \nu \Delta t \leq C/p^4$, while all nonlinearities, including the discretized subgrid term $-\textbf{D}\underline{\btau}$, are computed explicitly, e.g. by a second order extrapolation method (EX2), under the CFL restriction
\begin{equation}
\overline u_{\text{max}}\Delta t \leq C/p^2.
\end{equation}
 The implicit part is solved by a generalized block LU decomposition with a pressure correction algorithm \cite{perot93,deville02:_high,couzy95}. The overall order-in-time of the afore-presented numerical method is two.

\section{Subgrid modeling}
\subsection{General considerations}
The problem of subgrid modeling consists in taking into account the interaction between resolved and subgrid scales which is represented by the subgrid term $\bnabla\cdot\btau$ in the filtered momentum equation (\ref{eq:02-04}).

Following the terminology introduced by Sagaut \cite{sagaut03:_large}, two modeling strategies are defined. A first group of models, called structural, aims at making the best approximation of the tensor $\btau$ by reconstructing it formally from the resolved field $\overline{\mathbf{u}}$. The closure consists in finding a relation such that
\begin{equation}\label{eq:04-12}
\btau^{\mathrm{m}} = \mathcal{C}_{\btau}(\overline{\mathbf{u}}),
\end{equation}
where the upper index `$\mathrm{m}$' distinguishes the modeled from the exact subgrid tensor. This group of models does not require any foreknowledge about the nature of the interactions between resolved and subgrid scales. The second group, called functional, consists in modeling the action of subgrid scales on the resolved field $\obu$ using physical concepts and not at approximating the subgrid tensor $\btau$ itself, even if a subgrid tensor is explicitly constructed as for subgrid-viscosity models. Most of these models assume that the action of subgrid scales on resolved ones is essentially energetic, so that the balance of energy transfers between both scales categories is sufficient to describe the interaction.

The focus hereafter is on ADM which attempts to recover, at least partially, the original unfiltered fields, up to the grid level, by inverting the filtering operator applied to the Navier--Stokes equations. Following the idea of Winckelmans \etal\ \cite{winckelmans01:_asses_les} who coupled the ADM based on the van Cittert method and the Smagorinsky model \cite{smagorinsky63:_gener}, and Gullbrand and Chow \cite{gullbrand03} who proposed a dynamic version of the previous model, a new variant blending ADM and the dynamic mixed scale model introduced by Sagaut \cite{sagaut96:_numer} is proposed.

\subsection{Approximate deconvolution model}

The deconvolution approach aims at reconstructing the unfiltered fields from the filtered ones. The subgrid modes are not modeled but reconstructed using an \textit{ad hoc} mathematical procedure which falls in the structural modeling category. Writing formally the Navier--Stokes momentum equation \eqref{eq:02-04} as
\begin{equation}
\frac{\partial\bu}{\partial t}+\mathbf{f}(\bu) = \bm{0},
\end{equation}
the evolution equation of the filtered quantities becomes
\begin{equation}\label{eq:04-01}
\frac{\partial\obu}{\partial t}+\mathbf{f}(\obu) = [\mathbf{f},\mathcal{G}\star](\bu),
\end{equation}
where the convolution filter $\cG\star=(\mathcal{L}\circ\mathcal{P})\star$ embodies the LES filter $\cL \star$ and the projective grid filter $\cP \star$ \cite{winckelmans01:_asses_les,gullbrand03}, the latter being therefore implicitly accounted for in the sequel. It is important to note that the LES filter and the grid filter do not commute since the effect of the SEM discretization is not a spectral cutoff filter, unlike the case of spectral methods as reported by Gullbrand and Chow \cite{gullbrand03}. The subgrid commutator reads then
\begin{equation}
[\mathbf{f},\mathcal{G}\star](\bu)= \mathbf{f}(\cG\star \bu) - \cG \star \mathbf{f}(\bu) = \mathbf{f}(\obu ) - \overline{\mathbf{f}(\bu)},
\end{equation}
which is strictly equivalent to Eq. \eqref{eq:02-04} given
\begin{equation}\label{eq:04-51}
[\mathbf{f},\mathcal{G}\star](\bu) = -\bnabla\cdot\btau.
\end{equation}
The exact subgrid contribution appears as a function of the non-filtered field, which is not computed when performing a LES. This field being unknown, the idea is to approximate it using the following deconvolution procedure
\begin{equation}\label{eq:04-02}
\bu \simeq \bu^{*} = \mathcal{Q}_{N}\star\obu=(\mathcal{Q}_{N}\circ\mathcal{G})\star\bu =(\mathcal{Q}_{N}\circ\cL \circ \cP )\star\bu =(\mathcal{Q}_{N}\circ\cL )\star\hat{\bu} ,
\end{equation}
where $\hat{\bu}=\cP \star \bu$ is the grid-filtered velocity. The operator $\mathcal{Q}_{N}\star$ is an $N$th-order approximation of the inverse of the filter $\mathcal{L}\star$, since the grid filter is projective and therefore not invertible, such that
\begin{equation}
(\mathcal{Q}_{N}\circ\mathcal{L}) = \mathcal{I}+O(\overline\Delta^{N}),
\end{equation}
with $\cI\star$ the identity filtering operator and $\overline{\Delta}$ the filter cutoff length associated to $\cG\star$. Stolz and Adams proposed in \cite{stoltz99} an iterative deconvolution procedure based on the van Cittert method. If the filter $\cL \star$ has an inverse, it can be computed using the truncated van Cittert expansion series
\begin{equation}\label{eq:vancittert}
\mathcal{L}^{-1} \simeq \mathcal{Q}_{N} = \sum_{i=0}^{N}(\mathcal{I}-\mathcal{L})^{i},
\end{equation}
which is known to be convergent if
\begin{equation}\label{eq:04-77}
\left\|\mathcal{I}-\mathcal{L}\right\|\ll1.
\end{equation}
The deconvolution error induced by the approximation \eqref{eq:vancittert} can be represented by a filter $\cH_N\star$ defined by
\begin{equation} \label{eq:cH}
\cH_N = \cI - \cQ_N \circ \mathcal{L}.
\end{equation}
The subgrid term is then approximated as
\begin{equation}\label{eq:04-03}
[\mathbf{f},\mathcal{G}\star](\bu) \simeq [\mathbf{f},\mathcal{G}\star](\mathcal{Q}_{N}\star\obu) = [\mathbf{f},\mathcal{G}\star](\bua),
\end{equation}
and the model resulting from this approach is obtained by introducing Eq. (\ref{eq:04-03}) into the filtered Navier--Stokes momentum equation (\ref{eq:04-01})
\begin{equation}\label{eq:04-13}
\frac{\partial\obu}{\partial t}+\mathbf{f}(\obu) = [\mathbf{f},\mathcal{G}\star](\bu^{\ast}).
\end{equation}
Using once more approximation \eqref{eq:04-02} in Eq. \eqref{eq:04-13} implies $\mathbf{f}(\obu) = \mathbf{f}(\obu^{\ast})$ and leads to the formulation commonly used with ADM
\begin{equation}\label{eq:04-20}
\frac{\partial\obu}{\partial t}+\mathcal{G}\star\mathbf{f}(\bua) = 0.
\end{equation}
It is noteworthy that this latter formulation introduces the deconvolution error and the error related to the non-inversion of $\cP \star$ into the nonlinear advection term, thereby breaking the Galilean invariance \cite{speziale85:_galil}. Furthermore, the expression of the subgrid tensor of Bardina's scale similarity model \cite{bardina80:_improv} is not recovered from the deconvoluted formulation \eqref{eq:04-20} when {$\cQ_N=\cI$}, which is again due to the difference between the filtered and the deconvoluted velocities. Based on the previous comments, the filtered formulation \eqref{eq:04-13} appears to be the most general and therefore, all LES presented in the sequel rely on it. No numerical instabilities were observed using the formulation \eqref{eq:04-13} associated with our explicit treatment of the nonlinear terms, see Sec. \ref{sec:time-integration}.

\subsection{Coupling with a dynamic mixed scale model}
Coupling ADM with a subgrid-viscosity model can be formally achieved by adding a source term $\bs(\obu)$ to the right-hand side of Eq. (\ref{eq:04-13})
\begin{equation}\label{eq:04-16b}
\frac{\partial\obu}{\partial t}+\mathbf{f}(\obu) = [\mathbf{f},\mathcal{G}\star](\bu^{\ast}) + \mathbf{s}(\obu),
\end{equation}
where $\bs(\obu)$ is expressed in terms of the filtered rate-of-strain tensor $\overline{\mathbf{S}}$ by 
\begin{equation}
\mathbf{s}(\obu) = \bnabla\cdot(\nu_{\mathrm{sgs}}(\bnabla\obu+\bnabla\obu^{\mathrm{T}})) = \bnabla\cdot(2\nu_{\mathrm{sgs}}\overline{\mathbf{S}}),
\end{equation}
the superscript `$\mathrm{T}$' denoting the transpose operation and $\nu_{\mathrm{sgs}}$ the subgrid viscosity. For such functional models, only the deviatoric part of the subgrid stress is modeled. On the other hand, the ADM part $[\mathbf{f},\mathcal{G}\star](\bu^{\ast})$ includes both isotropic and deviatoric parts. Using such subgrid-viscosity model, the only unknown is the subgrid viscosity itself which implies a closure of the form
\begin{equation}\label{eq:04-17}
\nu_{\mathrm{sgs}}=\mathcal{C}_{\nu}(\obu).
\end{equation}

\subsubsection{Mixed scale model}
In the sequel, we focus on a subgrid-viscosity model proposed by Sagaut \cite{sagaut96:_numer} having a triple dependency on the large and small structures of the resolved field, and the filter cutoff length. With respect to the Smagorinsky model used by Winckelmans \etal\ \cite{winckelmans01:_asses_les}, the model proposed by Sagaut offers the advantage of automatically vanishing if subgrid scales are absent of the solution. This model, which makes up the one-parameter mixed scale family, is derived by taking a weighted geometric average of the models based on large scales and those based on the energy at cutoff. The closure is given by
\begin{equation}
\nu_{\mathrm{sgs}} = C_{\gamma}|\mathcal{F}(\obu)|^{\gamma}(\overline{q}_{\mathrm{c}})^{\frac{1-\gamma}{2}}\overline{\Delta}^{1+\gamma},
\end{equation}
where $C_{\gamma}$ and $\gamma$ are the subgrid-viscosity and mixed-scale constants, $\overline{q}_{\mathrm{c}}$ is the resolved kinetic energy at cutoff and
\begin{equation}
\mathcal{F}(\obu) = \mathbf{S}(\obu)=\overline{\mathbf{S}}\qquad\mathrm{or}\qquad
\mathcal{F}(\obu) = \bnabla\times\obu=\overline{\bm\omega}.
\end{equation}
The resolved kinetic energy at cutoff can be evaluated using the formula
\begin{equation}
\overline{q}_{\mathrm{c}}=\frac{1}{2}\ou_{\mathrm{c},i}\ou_{\mathrm{c},i},
\end{equation}
where the cutoff velocity field $\obu_{\mathrm{c}}$ represents the high-frequency part of the resolved field, defined using a second filter, referred to as test filter, designated by the \textit{tilde} symbol and associated with the cutoff length $\widetilde{\Delta}>\overline{\Delta}$
\begin{equation}
\obu_{\mathrm{c}}=\obu-\tobu.
\end{equation}
We note that for $\gamma\in[0,1]$, the subgrid viscosity is always defined. The constant $C_{\gamma}$ can be evaluated by theories of turbulence in the case of statistically homogeneous and isotropic turbulent flow
\begin{equation}\label{eq:04-47}
C_{\gamma}=C_{\mathrm{q}}^{1-\gamma}C_{\mathrm{s}}^{2\gamma},
\end{equation}
where the Smagorinsky constant $C_{\mathrm{s}} \simeq 0.18$ and $C_{\mathrm{q}} \simeq 0.20$. 

\subsubsection{Dynamic evaluation of the subgrid-viscosity constant}\label{sec:dynamic}
Theoretical values of the subgrid-viscosity constant cannot be used in our case because they are derived if the model is used without the ADM structural contribution, that is to model the whole subgrid tensor. In order to overcome this issue, we introduce a dynamic procedure of Germano--Lilly type to evaluate this parameter as a function of space and time. Such procedure completes the definition of the subgrid model based on the coupling of ADM with the dynamic mixed scale (DMS) model, called ADM-DMS in the sequel. This requires the introduction of the twice-filtered Navier--Stokes equations. Applying the test filter $\mathcal{T}\star$, represented by a $tilde$, to the filtered Navier--Stokes momentum equation (\ref{eq:04-01}) gives
\begin{equation}
\frac{\partial\tobu}{\partial t}+\mathbf{f}(\tobu) = [\mathbf{f},\mathcal{T}\star](\obu)+\mathcal{T}\star[\mathbf{f},\mathcal{G}\star](\bu),
\end{equation}
which can be recast in the form
\begin{equation}
\frac{\partial\tobu}{\partial t}+\mathbf{f}(\tobu) = - \bnabla\cdot(\mathbf{L}+\widetilde{\btau}) = - \bnabla\cdot\mathbf{T},
\end{equation}
where $\mathbf{T} = \mathbf{L}+\widetilde{\btau}$ is an expression of the Leibniz identity referred to as multiplicative Germano identity in the LES framework \cite{germano91}. The components of $\btau$ are given in Eq. \eqref{eq:deftau} and those of $\mathbf{L}$ by
\begin{equation}
L_{ij}=\widetilde{\ou_{i}\ou}_{j}-\tou_{i}\tou_{j},
\end{equation}
leading to the following expression for the subgrid tensor $\mathbf{T}$ corresponding to the twice-filtered Navier--Stokes equations
\begin{equation}
T_{ij} = \widetilde{\overline{u_{i}u}}_{j}-\tou_{i}\tou_{j}.
\end{equation}
The tensors corresponding to filtered and twice-filtered equations are modeled by blending ADM with the mixed scale model previously introduced. Assuming each subgrid tensor can be modeled using the same dynamic parameter $C_{\mathrm{d}}$ replacing the constant $C_{\gamma}$, which relies on the scale similarity hypothesis between test filter and primary filter cutoff lengths $\widetilde{{\Delta}}$ and $\overline{\Delta}$, we obtain
\begin{equation}\label{eq:04-16c}
\tau_{ij}^{\mathrm{m}}=\overline{\uia\uja}-\ouia\ \ouja+C_{\mathrm{d}}\beta_{ij},\qquad
\beta_{ij} = -2\overline\Delta^{1+\gamma}|\mathcal{F}(\obu)|^{\gamma}(\overline{q}_{\mathrm{c}})^{\frac{1-\gamma}{2}}\overline{S}_{ij},
\end{equation}
\begin{equation}\label{eq:04-17c}
T_{ij}^{\mathrm{m}}=\widetilde{\overline{\uia\uja}}-\touia\ \touja+C_{\mathrm{d}}\alpha_{ij},\qquad
\alpha_{ij} = -2\widetilde{\overline\Delta}^{1+\gamma}|\mathcal{F}(\tobu)|^{\gamma}(\widetilde{\overline{q}}_{c})^{\frac{1-\gamma}{2}}\widetilde{\overline{S}}_{ij},
\end{equation}
where $\bm\alpha$ and $\bm\beta$ are the subgrid-viscosity terms deprived of their constant. The parameter $C_{\mathrm{d}}$ is evaluated in order to minimize the residual
\begin{equation}\label{eq:04-28}
E_{ij}=L_{ij}-L_{ij}^{\mathrm{m}},
\end{equation}
where $\mathbf{L}^{\mathrm{m}}=\mathbf{T}^{\mathrm{m}}-\widetilde{\btau}^{\mathrm{m}}$. Using Eqs. (\ref{eq:04-16c})--(\ref{eq:04-17c}), Eq. (\ref{eq:04-28}) reads
\begin{equation}
E_{ij} = L_{ij}
-[(\widetilde{\overline{\uia\uja}}-\touia\ \touja+C_{\mathrm{d}}\alpha_{ij})
-(\widetilde{\overline{\uia\uja}}-\widetilde{\ouia\ \ouja}+\widetilde{C_{\mathrm{d}}\beta}_{ij})].
\end{equation}
Assuming $C_{\mathrm{d}}$ is constant over an interval at least equal to the test-filter cutoff length such that $\widetilde{C_{\mathrm{d}}\beta}_{ij}=C_{\mathrm{d}}\widetilde{\beta}_{ij}$, we have
\begin{equation}
E_{ij} = L_{ij}-(H_{ij}+C_{\mathrm{d}}m_{ij}),
\end{equation}
where
\begin{equation}
m_{ij}=\alpha_{ij}-\widetilde{\beta}_{ij}\qquad\mathrm{and}\qquad
H_{ij}=\widetilde{\ouia\ \ouja}-\touia\ \touja,
\end{equation}
which consists in a system of six independent equations leading to six possible different values of the constant. In a similar framework and in order to obtain a single value, Lilly \cite{lilly92:_german} proposed an evaluation based on a least-squares minimization of the form
\begin{equation}
\frac{\partial E_{ij}E_{ij}}{\partial C_{\mathrm{d}}}=0,
\end{equation}
leading to the solution of the following single scalar equation
\begin{equation}\label{eq:04-10}
C_{\mathrm{d}}=\frac{(L_{ij}-H_{ij})\ m_{ij}}{m_{ij}\ m_{ij}}.
\end{equation}
Smaller values than theoretical ones are expected for $\Cd$ using the previous dynamic procedure because of the small difference between the tensors $\mathbf{L}$ and $\mathbf{H}$, only induced by the deconvolution error. Indeed, the tensor $\mathbf{H}$ can be explicitly written as
\begin{equation}\label{eq:04-29}
H_{ij} = \widetilde{\overline{(\mathcal{Q}_{N}\star\ou_{i})}\ \overline{(\mathcal{Q}_{N}\star\ou_{j})}} - \widetilde{\overline{(\mathcal{Q}_{N}\star\ou_{i})}}\ \widetilde{\overline{(\mathcal{Q}_{N}\star\ou_{j})}},
\end{equation}
and if the deconvolution order $N\rightarrow\infty$, corresponding to $\mathcal{Q}_{N}\rightarrow\mathcal{G}^{-1}$ if the series \eqref{eq:vancittert} is convergent, one has
\begin{equation}\label{eq:04-88}
\lim_{N\rightarrow\infty}H_{ij}=L_{ij},
\end{equation}
which implies that the subgrid-viscosity term vanishes if exact deconvolution is performed up to the grid level. This behavior of the eddy-viscosity part of our model, when the deconvolution order tends to infinity is strictly equivalent to the one observed by Sagaut \etal\ \cite{sagaut00:_filter} and Stolz \etal\ \cite{stolz05:_high} using high-pass filtered subgrid-viscosity models. Furthermore the relaxation term introduced by Stolz \etal\ \cite{stolz01:compressible,stoltz01:incompressible} to stabilize their ADM-based LES has the same behavior in the infinite deconvolution order limit. The choice of the deconvolution order $N$ can be interpreted as a way of tuning the relative part taken by the subgrid-viscosity term which compensates the deconvolution error to minimize the difference between $\mathbf{L}$ and $\mathbf{L}^{\mathrm{m}}$, in a least-squares sense. In the limit of $N$ going to infinity, the modeled subgrid stress tensor defined in Eq. \eqref{eq:04-16c} reduces solely to its ADM contribution
\begin{equation}\label{eq:04-16d}
\tau_{ij}^{\mathrm{m}}=\overline{\hat{u}_i\hat{u}}_j-\overline{\hat{u}}_i\, \overline{\hat{u}}_j,
\end{equation}
where $\hat{\bu}=\mathcal{P}\star\bu$ is the grid-filtered velocity.

\subsection{Particular cases of ADM-DMS}\label{sec:particular-cases}

In this section we highlight two particular cases of ADM-DMS. The first one is the model proposed by Zang \etal\ \cite{zang93}, based on Bardina and Smagorinsky models with a dynamic evaluation of the subgrid-viscosity constant. The ADM-DMS expression of the subgrid tensor given by (\ref{eq:04-16c}) can be explicitly written as
\begin{equation}
\tau_{ij}^{m} = \overline{(\mathcal{Q}_{N}\star\ou_{i})(\mathcal{Q}_{N}\star\ou_{j})} - \overline{(\mathcal{Q}_{N}\star\ou_{i})}\ \overline{(\mathcal{Q}_{N}\star\ou_{j})}+C_{\mathrm{d}}\beta_{ij}. 
\end{equation}
Then choosing $\gamma=1$, $N=0$ and $\mathcal{F}(\obu)=\overline{\mathbf{S}}$ leads to
\begin{equation}
\tau_{ij}^{m} = \overline{\ou_{i}\ou}_{j}-\overline{\ou}_{i}\ \overline{\ou}_{j}+C_{\mathrm{d}}\beta_{ij},
\qquad \beta_{ij} = -2\overline\Delta^{2}|\overline{\mathbf{S}}|\overline{S}_{ij},
\end{equation}
which is the expression of the one-parameter dynamic mixed model. For $N=0$, the tensor $\mathbf{H}$ explicitly expressed by Eq. (\ref{eq:04-29}) reads
\begin{equation}
H_{ij} = \widetilde{\overline{\ou}_{i}\overline{\ou}}_{j} - \widetilde{\overline{\ou}}_{i}\widetilde{\overline{\ou}}_{j}.
\end{equation}

The second particular case of ADM-DMS is DMS, a dynamic version of the mixed scale model proposed by Sagaut \cite{sagaut96:_numer}. This model is formally obtained by imposing $\cQ_N=0$ in the developments of Sect. \ref{sec:dynamic}, which leads to $\mathbf{H}=\textbf{0}$ and to the following expression of the subgrid tensor
\begin{equation}
\tau_{ij}^{\mathrm{m}}=C_{\mathrm{d}}\beta_{ij},\qquad \beta_{ij} = -2\overline\Delta^{1+\gamma}|\mathcal{F}(\obu)|^{\gamma}(\overline{q}_{\mathrm{c}})^{\frac{1-\gamma}{2}}\overline{S}_{ij},
\end{equation}
with the dynamic parameter of DMS given by
\begin{equation}\label{eq:dyn2}
C_{\mathrm{d}}=\frac{L_{ij}\ m_{ij}}{m_{ij}\ m_{ij}}.
\end{equation}
Without the ADM contribution, higher values of the dynamic parameter are expected since the difference between $\mathbf{L}$ and $\mathbf{H}$ occurring in Eq. \eqref{eq:04-10} disappears in Eq. \eqref{eq:dyn2}. This phenomenon is in direct relation with the fact that the subgrid viscosity term is used to model the whole subgrid tensor in this particular case.

\section{Filtering}
Filtering techniques suited to SEM\ and LES must preserve $C^{0}$-continuity of the filtered variables across spectral elements and be applicable at the element level. In the sequel, we present a filter satisfying these constraints which is based on spectral techniques ensuring the element-level filtering property. The filtering operation is performed by applying a given transfer function in a modal basis. Depending on this transfer function, this filter may not be projective, therefore ensuring its invertibility which is a key feature needed by the deconvolution procedure. Hence, we will focus on the choice of the transfer function to fulfill this constraint.

\subsection{Description of the filter}
The modal basis introduced in the $p$-version of finite elements and first used by Boyd \cite{boyd98:_two_cheby_legen} as filtering technique is presented in its one-dimensional version, the extension to three dimensions being straightforward by tensor product. It is built up on the reference parent element $\hat\Omega=[-1,1]$ of the SEM\ as
\begin{equation}
\begin{array}{ll}
\displaystyle{\phi_{0}=\frac{1-\xi}{2}},\qquad \displaystyle{\phi_{1}=\frac{1+\xi}{2}}, \\[3mm]
\phi_{j}=L_{j}(\xi)-L_{j-2}(\xi), & 2 \leq j \leq p,
\end{array}
\end{equation}
where $L_j$ is the Legendre polynomial of degree $j$. Unlike the Lagrange--Legendre nodal basis used in our spectral element calculations, this modal basis forms a hierarchical set of polynomials allowing to define in an explicit and straightforward manner a low-pass filtering procedure. Any variable $v$ can be expressed in this basis by the relation
\begin{equation}\label{eq:05-01}
v(\xi)=\sum_{j=0}^{p}\breve{v}_{j}\phi_{j}(\xi), \qquad \xi\in\hat\Omega,
\end{equation}
which in matrix notation reads
\begin{equation}
\mathbf{v}=\mathbf{\Phi}\breve{\mathbf{v}},
\end{equation}
where
\begin{equation}
\Phi_{ij}=\phi_{j}(\xi_{i}).
\end{equation}
The filtering operation is performed in the spectral modal space through a diagonal matrix $\mathbf{K}$ whose components are chosen in order to fulfill the required properties of the filter. The filtering process for a one-dimensional problem is expressed by
\begin{equation}
\overline{\mathbf{v}}=\mathbf{\Phi}\mathbf{K}\mathbf{\Phi}^{-1}\mathbf{v}=\mathbf{G}\mathbf{v}.
\end{equation}

\subsection{Transfer function}
\begin{figure}[htb]
        \centering
                \includegraphics[width=0.80\textwidth]{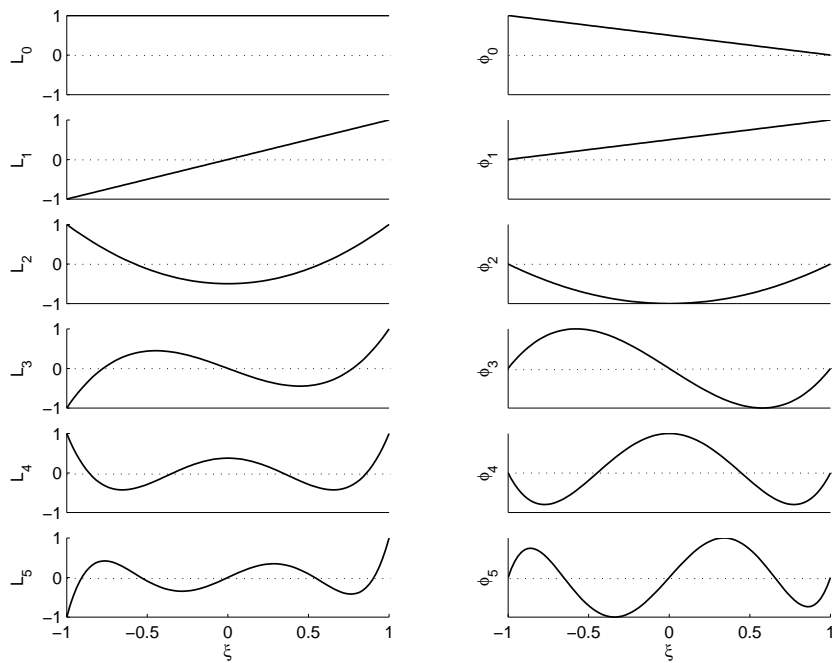}
        \caption{Bases associated with the filtering operation, shown for polynomial order $p=5$ on the reference parent element $\hat\Omega = [-1,1]$. The Legendre polynomials (left column), the modal polynomials (right column).}
        \label{fig:05-01}
\end{figure}

\begin{figure}[htb]
        \centering
                \includegraphics[width=0.49\textwidth]{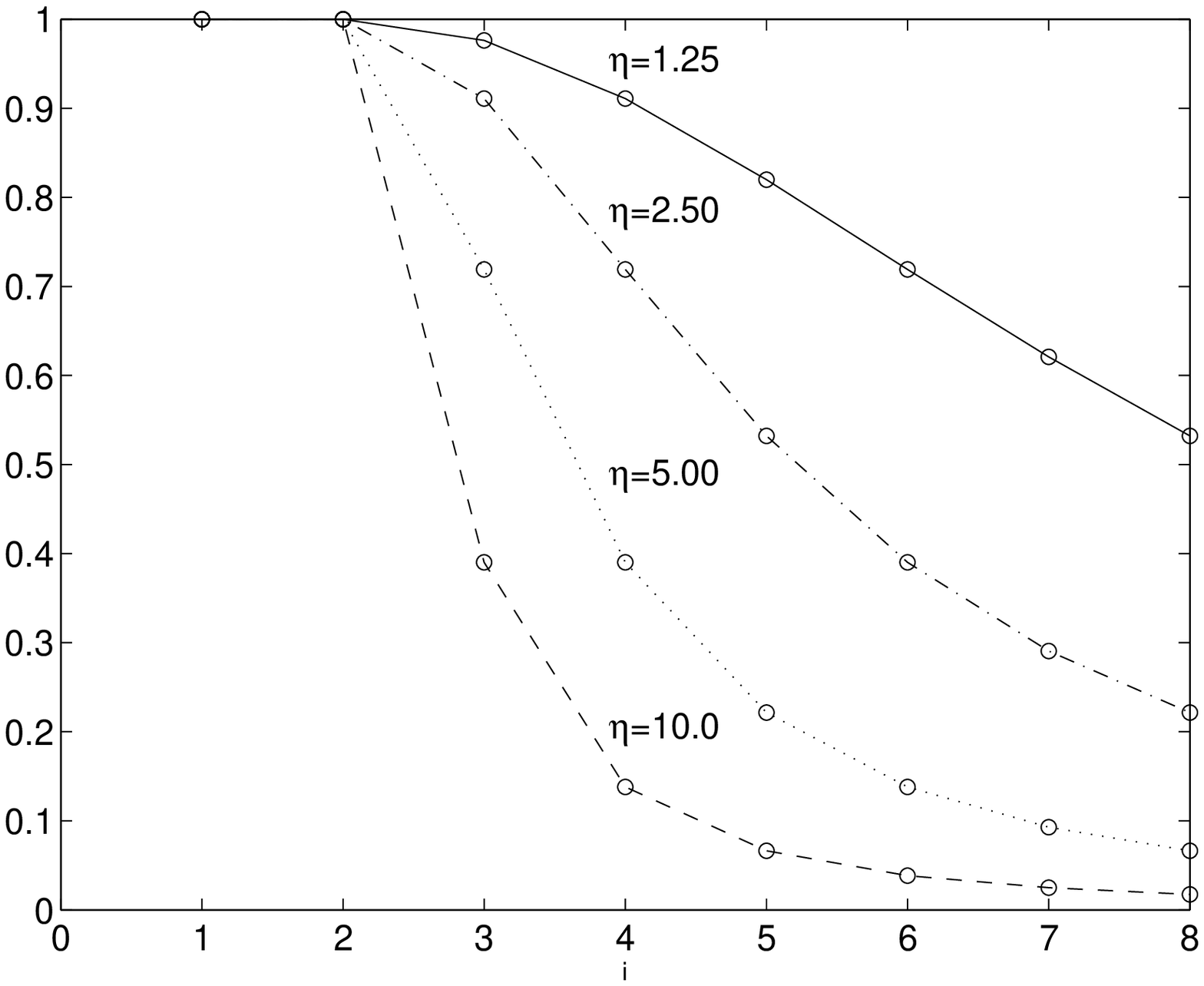}
                \includegraphics[width=0.49\textwidth]{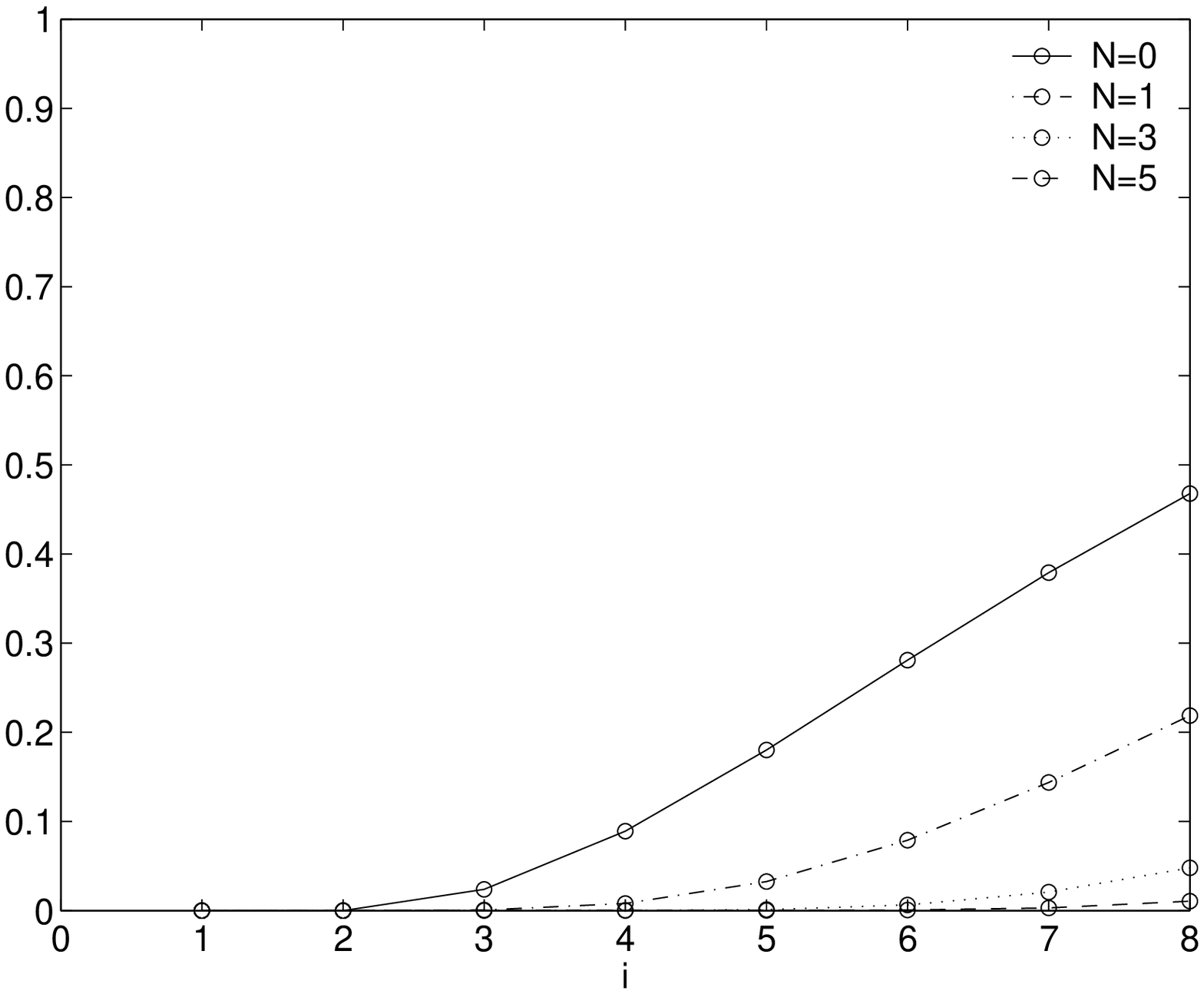}
        \caption{Transfer functions associated with $\mathcal{G}\star$ for different values of the filtering rate $\eta=1.25,\ 2.50,\ 5.00,\ 10.0$ (left) and, $\mathcal{H}\star$ for different values of the deconvolution order $N=0,\ 1,\ 3,\ 5$ with $\eta=1.25$ (right). Polynomial order $p=8$.}
        \label{fig:05-02}
\end{figure}

$C^{0}$-continuity, conservation of constants, invertibility and low-pass filtering are obtained by properly choosing the transfer function represented by the diagonal transfer matrix $\mathbf{K}$. Imposing all these requirements to the filter could seem like an intractable issue but appears feasible when visualizing the modal basis functions presented in \cite{blackburn03:_spect} and reported in Fig. \ref{fig:05-01}. As the filter acts in another basis than the one used for our spectral element calculations, $C^{0}$-continuity is preserved if the boundaries of the elements are not affected by the filtering procedure. One can notice that the only shape functions having non-zero values at the element boundaries are $\phi_{0}$ and $\phi_{1}$, while $\phi_j,\ j\geq 2$ are bubble functions. The functions $\phi_{0}$ and $\phi_{1}$ are responsible for imposing the non-zero values on element edges. Therefore, the transfer function coefficients must satisfy
\begin{equation}\label{eq:05-03}
K_{ij} = \delta_{ij}, \qquad i,\,j\leq1,
\end{equation}
with $\delta_{ij}$ the Kronecker operator. If $K_{ij}$ verifies (\ref{eq:05-03}), the constants are conserved after filtering because they are expressed as a linear combination of $\phi_{0}$ and $\phi_{1}$. The modal filter is not projective if all diagonal coefficients $K_{ii}$ are non-zeros. The last required property is to perform low-pass filtering in frequency. As this modal basis forms a hierarchical set of polynomials, low-pass filtering is done by damping the high-degree polynomial contributions. The transfer matrix is expressed by
\begin{equation}
K_{ij}=\delta_{ij}\cK(i),
\end{equation}
with the continuous transfer function
\begin{equation}\label{eq:05-02}
\cK(k)=\frac{1}{1+\left(\eta\frac{\max(0,k-1)}{p}\right)^{2}}, \qquad \eta \geq 0,
\end{equation}
where $\eta$ is referred to as filtering rate (Fig. \ref{fig:05-02}). The transfer function is such that the filter verifies all the required properties previously described. The cutoff frequency $\overline{k}$ is arbitrarily defined by $\cK(\overline{k})=1/2$. Such filtering technique has already been used by Blackburn and Schmidt for the LES of channel flow using SEM\  \cite{blackburn03:_spect}. In the present work, the transfer function given by Eq. \eqref{eq:05-02} and depicted on Fig. \ref{fig:05-02} ensures the invertibility of the filter contrary to \cite{blackburn03:_spect}. Moreover, the shape of the transfer function in Fig. \ref{fig:05-02} is similar to the one classically used by Stolz \etal\ \cite{stoltz01:incompressible,stolz01:compressible}. However, in the SEM framework the transfer function is defined element by element in the spectral modal space which prevents from a direct comparison with the discrete filter implemented by Stolz \etal\ in \cite{stoltz01:incompressible,stolz01:compressible}.

\subsection{Filter cutoff length}
The subgrid-viscosity term of ADM-DMS makes explicitly use of the filter cutoff length which needs to be defined. For a one-dimensional problem, e.g. in the $x$-direction, using the SEM, a common choice of filter width \cite{bouffanais05:_large,bouffanais06:_large} is
\begin{equation}
\overline{\Delta}_{x}=\frac{\hat{\Delta}_{x}}{p_{x,\mathrm{c}}},
\end{equation}
where $\hat{\Delta}_{x}$ is the element size and $p_{x,\mathrm{c}}$ the highest polynomial degree in the spectral decomposition (\ref{eq:05-01}) that is the closest to the cutoff frequency $\overline{k}$
\begin{equation}
p_{x,\mathrm{c}}=\max(i),\qquad\mathrm{such\ that\ }i\leq\overline{k},\qquad i=0,\ldots,p.
\end{equation}
We notice that the filter length decreases when the element is refined and the polynomial degree augmented. The straightforward three-dimensional extension for problems with rectilinear spectral elements is
\begin{equation}
\overline{\Delta}(x,y,z)=(\overline{\Delta}_{x}(x)\overline{\Delta}_{y}(y)\overline{\Delta}_{z}(z))^{1/3}=
\left(\frac{\hat{\Delta}_{x}}{p_{x,\mathrm{c}}}\frac{\hat{\Delta}_{y}}{p_{y,\mathrm{c}}}\frac{\hat{\Delta}_{z}}{p_{z,\mathrm{c}}}\right)^{1/3}.
\end{equation}

\subsection{Filtering operators related to ADM}
The filtering operators $\mathcal{Q}_{N}\star$ and $\mathcal{H}_N\star$ are defined with respect to $\mathcal{G}\star$, see Eq. \eqref{eq:vancittert} and \eqref{eq:cH} respectively and explicitly depend on the deconvolution order $N$. By representing in Figure \ref{fig:05-02} the transfer function associated with $\mathcal{H}_N\star$, one can observe that the deconvolution error is important at the end of the modal spectrum, so that $\mathcal{H}_N\star$ constitutes a high-pass filter. When increasing the deconvolution order $N$, the transfer function associated with the filter $\cH_N\star$ diminishes, showing the increasing accuracy of the approximate deconvolution procedure.

\section{LES of the lid-driven cubical cavity flow}
\subsection{General considerations}

\begin{figure}[htb!]
        %\centering
  \hspace{-2cm}
        \input{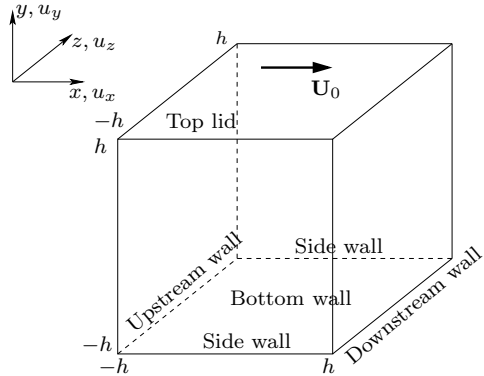}
                \caption{Lid-driven cubical cavity. Geometry and definitions.}
        \label{fig:06-01}
\end{figure}

The different LES presented hereafter refer to the flow in a lid-driven cubical cavity performed at Reynolds number of 12'000. The flow domain $\bm\Omega$ consists in a cubical cavity such that $\bm\Omega=(-h,h)^3$, the axis origin being assigned at the center of the cavity (Fig. \ref{fig:06-01}). The flow is driven by imposing a non-zero velocity parallel to the $x$-axis on the ``top'' wall. On the other walls, no-slip conditions are imposed. The moving wall will be referred to as the lid while the faces normal to the $z$-axis will be referred to as side walls. The upstream and downstream walls are normal to the $x$-axis and characterized by their relative position with respect to the lid motion. The remaining face parallel to the lid is called bottom wall. As far as the velocity imposed on the lid is concerned, the unit velocity induces severe discontinuities along the top edges. In order to avoid these defects, the imposed velocity on the lid is given by the polynomial expression
\begin{equation}
u_{x}(x,h,z)=U_{0}(1-(x/h)^{n})^{2}(1-(z/h)^{n})^{2}, u_{y}=u_{z}=0,
\end{equation}
where $U_{0}$ is a constant. The Reynolds number is defined using the maximum velocity $U_{0}$
\begin{equation}
\mathrm{Re}=\frac{2hU_{0}}{\nu}.
\end{equation}

Although the geometry is very simple, the flow presents complex physical phenomena \cite{leriche00:_direc,bouffanais06:_large}, no direction of homogeneity and a large variety of flow conditions. For such Reynolds numbers, the flow over most of the domain is laminar and turbulence develops near the cavity walls. Its main feature is a large scale recirculation which spans the cavity in the $z$-direction. Aside this large flow structure, the relatively high momentum fluid near the lid is deviated by the downstream wall into a down flowing nonparallel wall jet which separates ahead of the bottom wall. A region of high pressure and dissipation located at the top of downstream wall results from this deviation. The energy resulting from the impingement of the separated layer against the bottom wall is lost to turbulence and partly recovered by an emerging wall jet near the upstream wall where the flow slows down and relaminarizes during the fluid rise. The flow is also characterized by multiple counter-rotating recirculating regions at the corners and edges of the cavity.
\begin{table}[!ht]
        \centering\small{
                \begin{tabular}{lccccc}
                        & Time-step & Lid vel. & Int. time & Nb. elements & Polynomial degree \\
                        & $h/U_{0}$ & $n$ & $h/U_{0}$ & $(E_x,E_y,E_z)$ & $(p_x,p_y,p_z)$ \\\hline\hline
                        DNS & 0.0025 & 18 & 1'000 & $(1,1,1)$ & $(128,128,128)$ \\
                        LES & 0.0020 & 18 & 80    & $(8,8,8)$ & $(8,8,8)$       \\\hline
                \end{tabular}}
        \caption{Numerical and physical parameters of the DNS \cite{leriche00:_direc} and LES.}
        \label{tab:07-01}
\end{table}

\begin{figure}[!ht]
        \centering
                \includegraphics[width=0.40\textwidth]{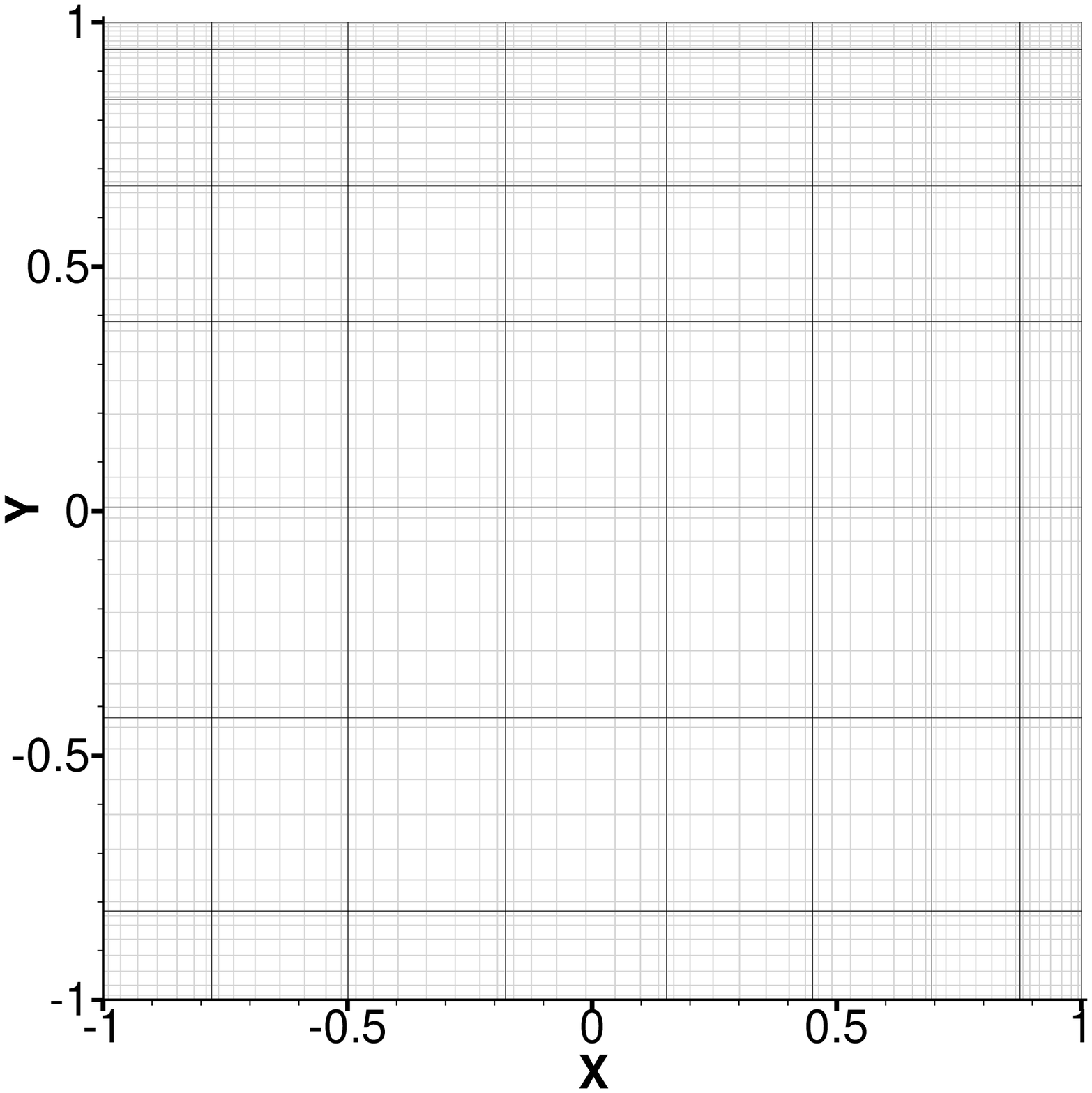}
                \caption{Spectral element grid in any plane normal to the $z$-direction.}
        \label{fig:07-02b}
\end{figure}

The physical and numerical parameters of the DNS and the LES are gathered in Table \ref{tab:07-01}. The DNS constitutes the reference solution and was obtained with a Chebyshev collocation method on a grid composed of $129$ collocation points in each spatial direction \cite{leriche00:_direc}. For LES, the spectral elements are unevenly distributed (Fig. \ref{fig:07-02b}) in order to resolve the boundary layers along the lid and the downstream wall. The spatial discretization has $E_{x}=E_{y}=E_{z}=8$ elements in the three space directions with $p_{x}=p_{y}=p_{z}=8$ polynomial degree, equivalent to $65^{3}$ grid points in total. The mesh used for LES has therefore twice less points per space direction than the DNS grid of Leriche and Gavrilakis but it is important to note that to achieve a DNS using the SEM would require more than $129^3$ grid points due to the lower order of the SEM as compared to the Chebyshev collocation method \cite{deville02:_high}. The space discretization is strictly equivalent to the one used for the LES reported by Zang \etal\ \cite{zang93} for a lower Reynolds number of 10'000. One should notice that the time-step for LES is slightly smaller than for the DNS which is due to different CFL constraints for the two different numerical schemes used, namely SEM\ and Chebyshev collocation. 

The mixed scales constant is set to $\gamma=0.5$ in order to have the triple dependency on the large and small structures of the resolved field as a function of the filter cutoff length. Furthermore, the ratio between both filtering rates $\eta_\cT$ and $\eta_\cG$ in Eq. \eqref{eq:05-02}, corresponding to the test and primary filters $\cT\star$ and $\cG\star$ respectively, is taken equal to two leading to a ratio of the filter cutoff lengths $\widetilde{\Delta}/\overline{\Delta}=7/4$. The parameters chosen for all LES analyzed hereafter are summarized in Table \ref{tab:07-02}. The choice of the deconvolution order is based on the observations of Stolz \etal \cite{stoltz01:incompressible,stolz01:compressible} and, Gullbrand and Chow \cite{gullbrand03} who found that the value $N=5$ for the deconvolution order is a good compromise between the precision in the approximate deconvolution and the computational cost induced by higher $N$ in the van Cittert expansion series. This choice is further justified by the analysis of the approximate deconvolution error developed in Sec. \ref{sec:validation-ADM}.

A LES based on ADM-DMS with the same parameters as the ones in Table \ref{tab:07-02}, except for $\cF=\overline{\bomega}$ has been carried out and has provided results extremely close to those from ADM-DMS with $\cF=\overline{\bS}$. Consequently, ADM-DMS results for the case $\cF=\overline{\bS}$ are the only ones reported in this article. A LES based on DMS, see Sec. \ref{sec:particular-cases}, with the same parameters as ADM-DMS for its dynamic mixed scale part, is also presented and compared to ADM-DMS in order to identify the improvement induced by coupling ADM with DMS.

\begin{table}[!ht]
        \centering\small{
                \begin{tabular}{cccccc}
                LES model & $N$ & $\eta_{\mathcal{G}}$ & $\eta_{\mathcal{T}}$ & $\gamma$ & $\cF$\\\hline\hline
                ADM-DMS & 5 & 1.25 & 2.50 & 0.50 & $\overline{\bS}$ \\
                DMS & - & 1.25 & 2.50 & 0.50 & $\overline{\bS}$ \\
                \hline
                \end{tabular}}
                \caption{Models parameters for both ADM-DMS and DMS.}
                \label{tab:07-02}
\end{table}

The different LES are all started from the same initial condition, namely an instantaneous velocity field obtained from the DNS in the statistically-steady range and re-interpolated onto the spectral element grid. The projective filter due to this re-interpolation induces the unrecoverable loss of the subgrid scales.

\begin{figure}[!ht]
        \centering
                \includegraphics[width=0.32\textwidth]{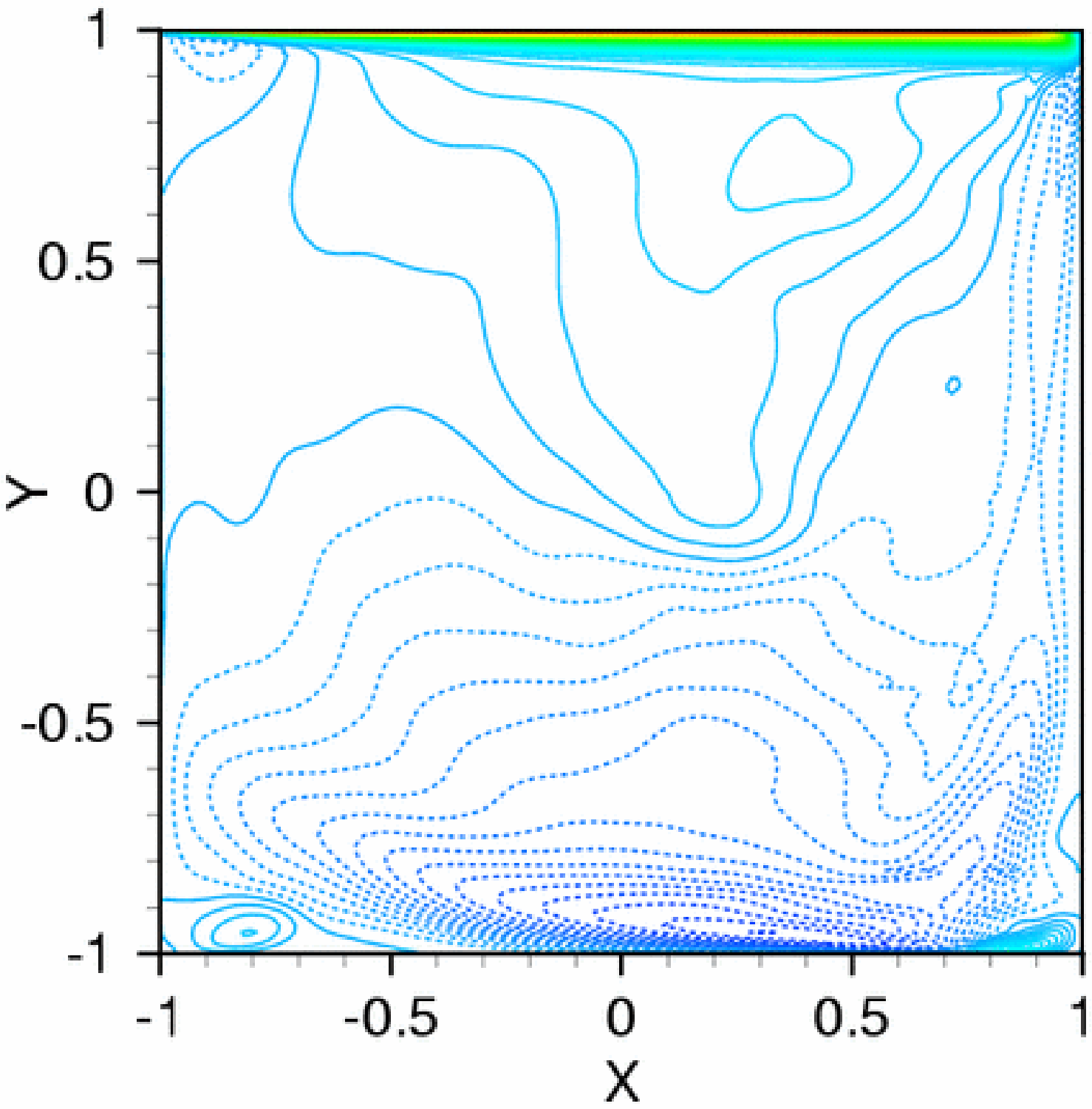}
                \includegraphics[width=0.32\textwidth]{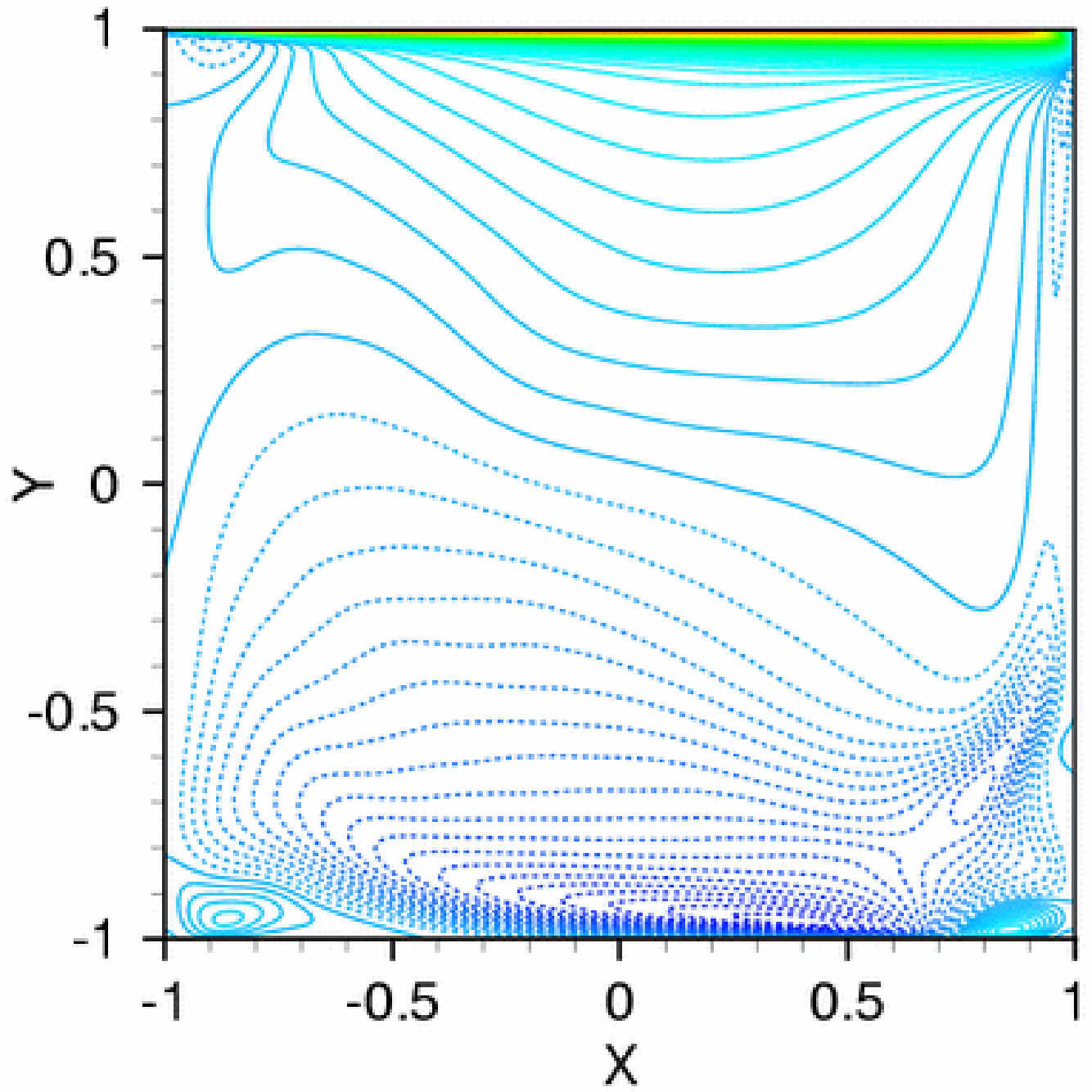} \\
                \includegraphics[width=0.32\textwidth]{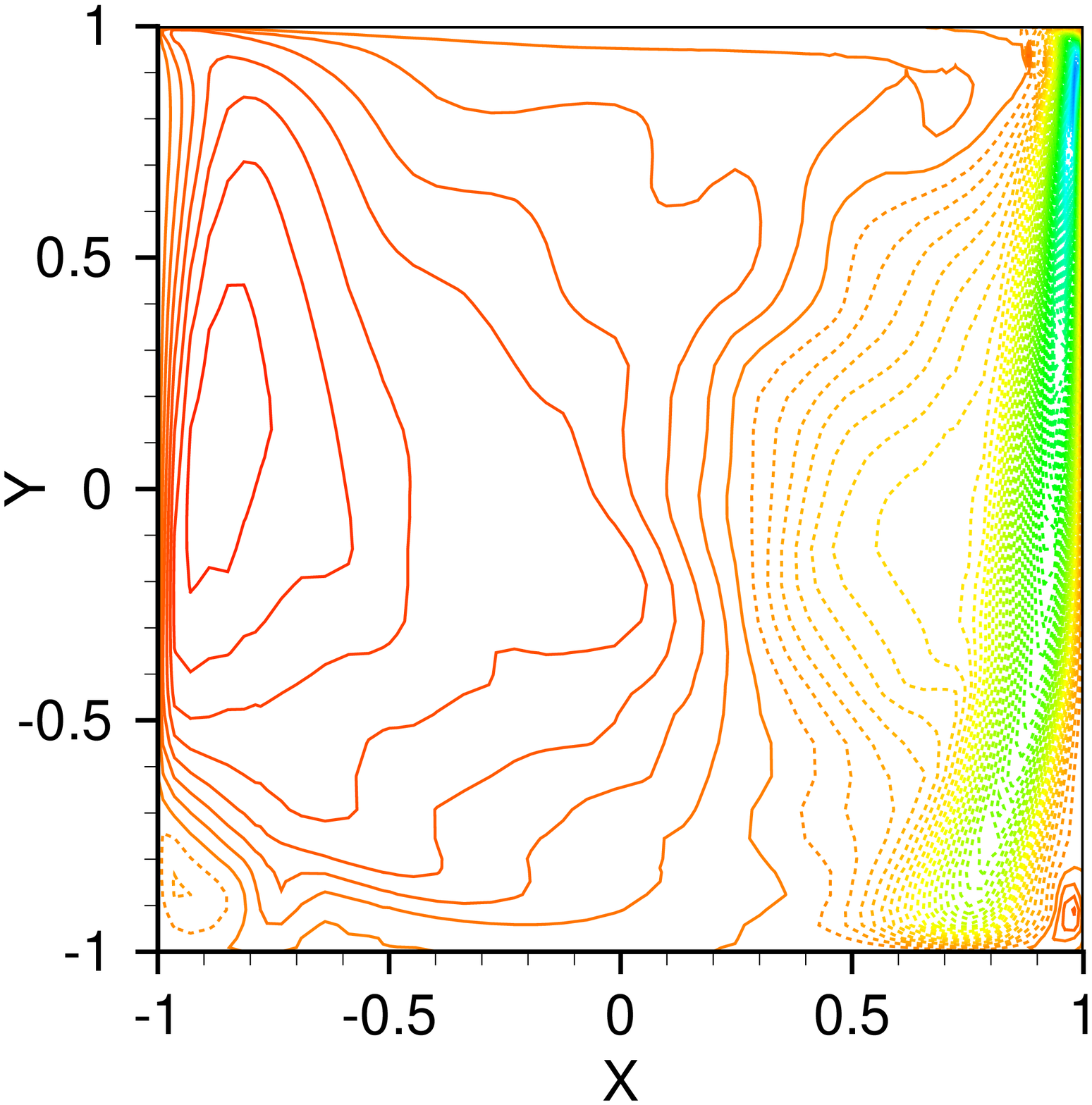}
                \includegraphics[width=0.32\textwidth]{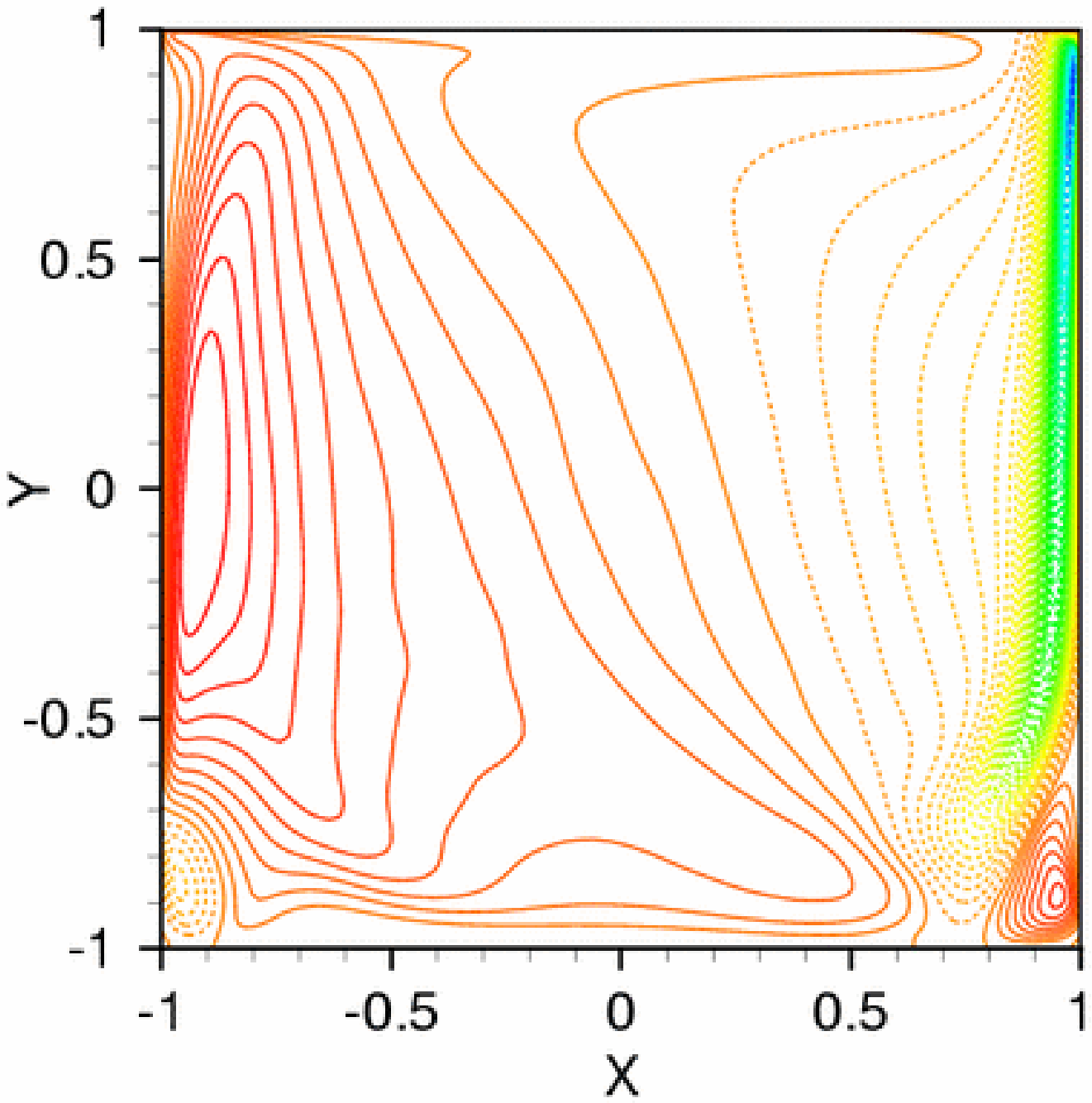}
                \caption{In the mid-plane $z/h=0$, UDNS (left column) and DNS (right column). Top row: contours of the $x$-component of the average resolved velocity field from $-0.2$ to $1$ by increments of $0.01$. Bottom row: contours of the $y$-component of the average resolved velocity field from $-0.7$ to $0.1$ by increments of $0.01$. Color scale from blue to red. Dashed contours correspond to negative levels. Levels in $U_0$ units.}
                \label{fig:07-06}
\end{figure}

In order to verify that our mesh is coarse enough and does not resolve all scales of the flow, a DNS of the lid-driven cubical cavity flow was performed with SEM\ and with exactly the same physical and numerical parameters as the ones reported in Table \ref{tab:07-01}. One can observe on Fig. \ref{fig:07-06} that this under-resolved DNS (UDNS) is totally inoperative in the particular context of this simulation. Even first-order statistics are far from being well predicted, not to mention second-order ones. These results allow us to confirm the sufficient under-resolution of the flow using the $65^3$ SEM mesh.

\begin{figure}[htb!]
       \centering
        \input{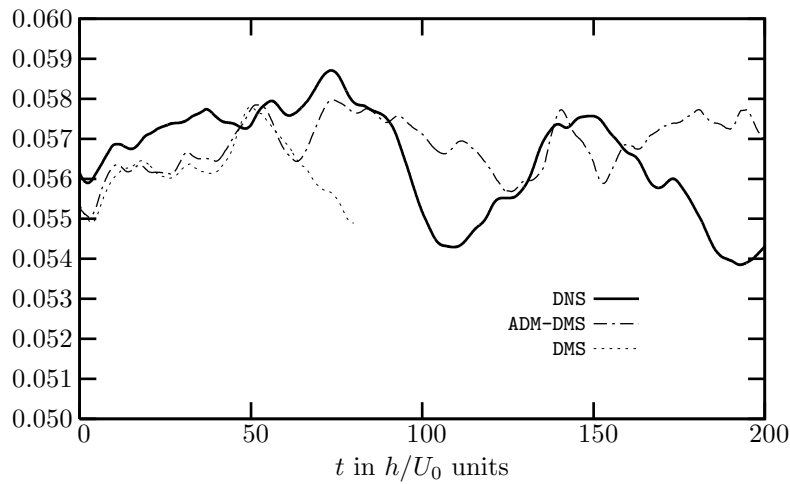}
                \caption{Total resolved kinetic energy $Q(\overline{\bu})$ in $U_0^2h^3$ units with respect to the time $t$ in $h/U_0$ units and, for the DNS, ADM-DMS and DMS (limited to 80 time units).}
        \label{fig:KineticEnergy}
\end{figure}

We assume that a statistically-steady state is reached and time averaging will be taken as ensemble averaging. For any variable $v$, the Reynolds statistical decomposition
\begin{equation}
v = \langle v\rangle+v^{\circ}
\end{equation}
introduces the time-averaged value denoted into brackets $\langle v\rangle$ and its fluctuating part $v^{\circ}$. It is noteworthy reminding here the difference between the filter splitting $v=\overline{v}+v^\prime$ and the Reynolds decomposition. As the initial condition of all LES is the same DNS instantaneous velocity field taken from the statistically-steady-state range, it is reasonable to also assume that LES will reach a statistically-steady state very quickly, if subgrid modeling is efficient \cite{bouffanais06:_large}. These assumptions are easily verified by evaluating the total kinetic energy of the resolved field
\begin{equation}
Q(\overline{\bu}) = \frac{1}{2}\int_{\Omega}\overline{u}_{i}\overline{u}_{i}\, \mathrm{d}\Omega,
\end{equation}
which is expected to evolve within a relatively small fluctuation range. For figure \ref{fig:KineticEnergy}, the results reported for ADM-DMS correspond to a longer dynamic range of $200\,h/U_0$ time units. However, all the statistical results presented hereafter for both ADM-DMS and DMS are limited to the first $80\,h/U_0$ time units. The time histories of $Q(\overline{\bu})$ presented in Fig. \ref{fig:KineticEnergy} for ADM-DMS and DMS show an evolution within the same fluctuation range as the DNS and around the average value of the total kinetic energy predicted by the DNS. As reported by Bouffanais \etal\ \cite{bouffanais06:_large} using a dynamic Smagorinsky model, which is a particular case of the present DMS over $800\, h/U_0$ time units, further confirms the evolution of $Q(\overline{\bu})$ for DMS in the long run.

\begin{figure}[htb!]
       \centering
        \input{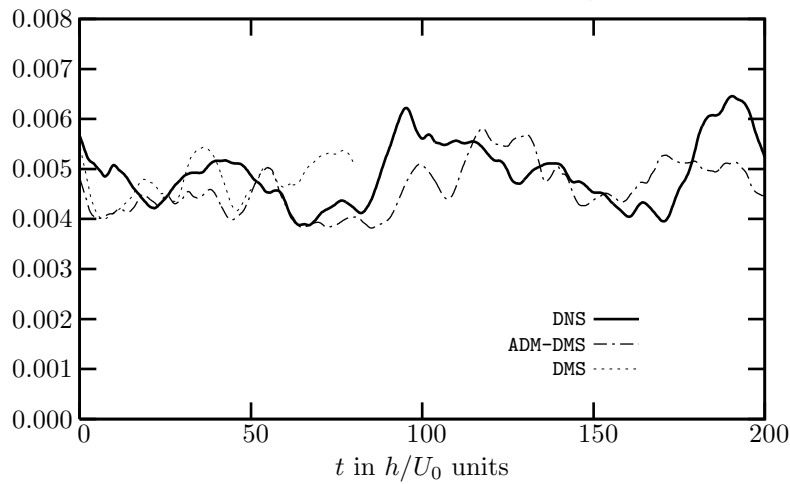}
                \caption{Fluctuating resolved kinetic energy $Q(\overline{\bu}^\circ)$ in $U_0^2h^3$ units with respect to the time $t$ in $h/U_0$ units and, for the DNS, ADM-DMS and DMS (limited to 80 time units).}
        \label{fig:FluctuatingKineticEnergy}
\end{figure}

Additionally the time histories of the kinetic energy of the fluctuating resolved field $Q(\overline{\bu}^\circ)$ presented on Fig. \ref{fig:FluctuatingKineticEnergy} is also evolving in the same fluctuation range as the DNS. The results on Fig. \ref{fig:KineticEnergy} and Fig. \ref{fig:FluctuatingKineticEnergy} for both $Q(\overline{\bu})$ and $Q(\overline{\bu}^\circ)$ allow to conclude that the turbulent flow reaches a statistically-steady state extremely quickly. No transient can be clearly identified in this case. The same conclusion is made by Bouffanais \etal\ \cite{bouffanais06:_large} for LES using more classical subgrid models.

Since the scale separation used for LES leads to the removal of subgrid scales mainly responsible for the energy dissipation, the subgrid model has to take into account this phenomenon. The flow in the cavity is confined and recirculating so that the same fluid is conserved inside the cavity. Moreover, kinetic energy is constantly provided to it by viscous diffusion. Hence, integral energy quantities over the flow domain such as $Q(\overline{\bu})$ and $Q(\overline{\bu}^\circ)$ are a direct indication of any under- or over-dissipative character of the subgrid model, keeping in mind the very low numerical dissipation and dispersion of the SEM. The results obtained for $Q(\overline{\bu})$ using ADM-DMS and DMS clearly show that the energy balance is achieved when using these models in this context.

\subsection{Validation of the approximate deconvolution procedure}\label{sec:validation-ADM}

The first step towards a complete validation of the ADM-DMS model, resides in investigating the accuracy of the deconvolution procedure based on the van Cittert method, with respect to the deconvolution order $N$. For this purpose, we define the relative error in $\mathrm{L}^{2}$-norm between a non-filtered DNS velocity field, extracted from the DNS database of Leriche and Gavrilakis \cite{leriche00:_direc}, and its deconvoluted counterpart $\cQ_N\star\obu$
\begin{equation}
e_{\bu}=\frac{\left\|\bu-\mathcal{Q}_{N}\star\obu\right\|_{\mathrm{L}^{2}(\bm\Omega)}}{\left\|\bu\right\|_{\mathrm{L}^{2}(\bm\Omega)}}.
\end{equation}

\begin{figure}[htb!]
        \centering
        \input{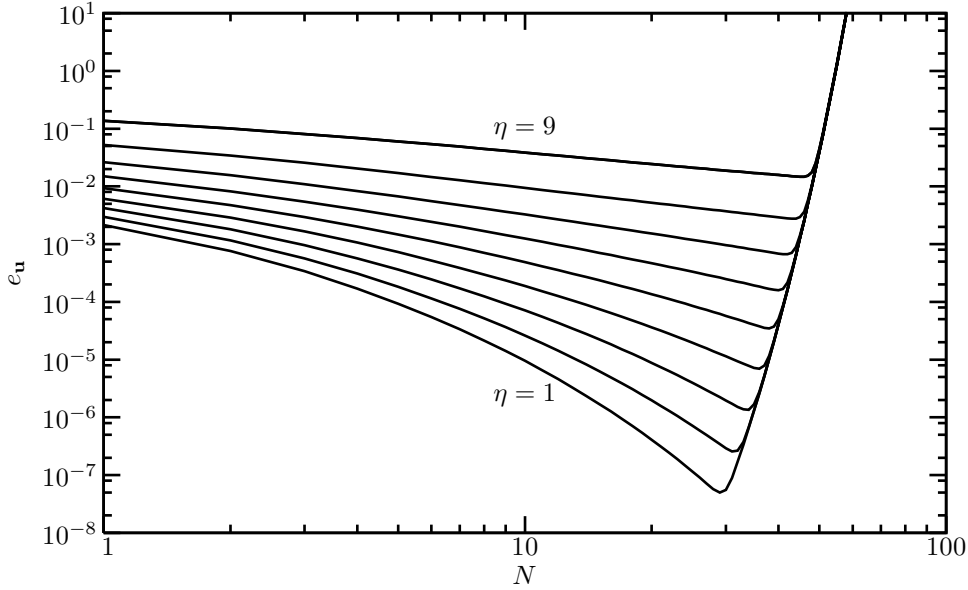}
        \caption{Parametric analysis of the deconvolution error $e_\bu$ with respect to the deconvolution order $N$. The filtering rate $\eta$ is increased from 1 to 9 with unit increments.}
        \label{fig:06-02}
\end{figure}

Figure \ref{fig:06-02} displays the parametric analysis of the relative error with respect to the deconvolution order, with the filtering rate $\eta_{\mathcal{G}}$ as parameter. One can notice that the van Cittert expansion series is convergent and the error increases with the filtering rate $\eta_{\mathcal{G}}$. In practical tests the deconvolution order must be set lower to $30$ in order to avoid having binomial coefficients of very high values which would inevitably induce precision errors. This numerical issue justifies the ``apparent divergence'' of the approximate deconvolution procedure for all filtering rate observed in Fig. \ref{fig:06-02} for large values of $N$. The filtering rate is increased from 1 to 9 by unit increments showing that the deconvolution error is larger with higher values of the filtering rate, which corresponds to the expected result. One can also notice that the error growth in the ``apparent divergence'' occurs earlier with lower filtering rates. It is very interesting to note that the error analysis is being performed using a velocity field corresponding to a turbulent flow including laminar regions. The resulting deconvolution error is clearly higher than the one obtained with a smooth analytically-defined field.

\subsection{\textit{A posteriori} validations}

In this section, results of the LES are compared with the available reference results by analyzing first- and second-order statistics. The measurements reported by Leriche and Gavrilakis \cite{leriche00:_direc} were taken in the mid-plane $z/h=0$, which is the statistical symmetry plane of the flow domain. For the sets of DNS data, the total velocity field is considered whereas in the case of LES, only its resolved part is presented. In consequence, the statistical moments computed from the resolved field cannot be equal to those computed from the DNS. One solution to overcome this issue would have been to apply the same filtering as is used for the scale separation to the reference solution \cite{moeng88:_spect}. We refer the reader to the monograph by Sagaut \cite{sagaut03:_large} for more details. The statistics for all LES and UDNS are based on a sampling approximately 10 times smaller than the one of the original DNS, but about twice longer than the one of the LES of Zang \etal \cite{zang93}; more precisely 400 samples are collected over $80\,h/U_0$ time units. The original reference DNS statistics were performed using 4'000 samples extracted over an integration range of 1'000$\,h/U_{0}$. Therefore, the LES statistics are not expected to be identical to the reference ones, especially the second-order ones.

\begin{figure}[!ht]
        \centering
                \includegraphics[width=0.32\textwidth]{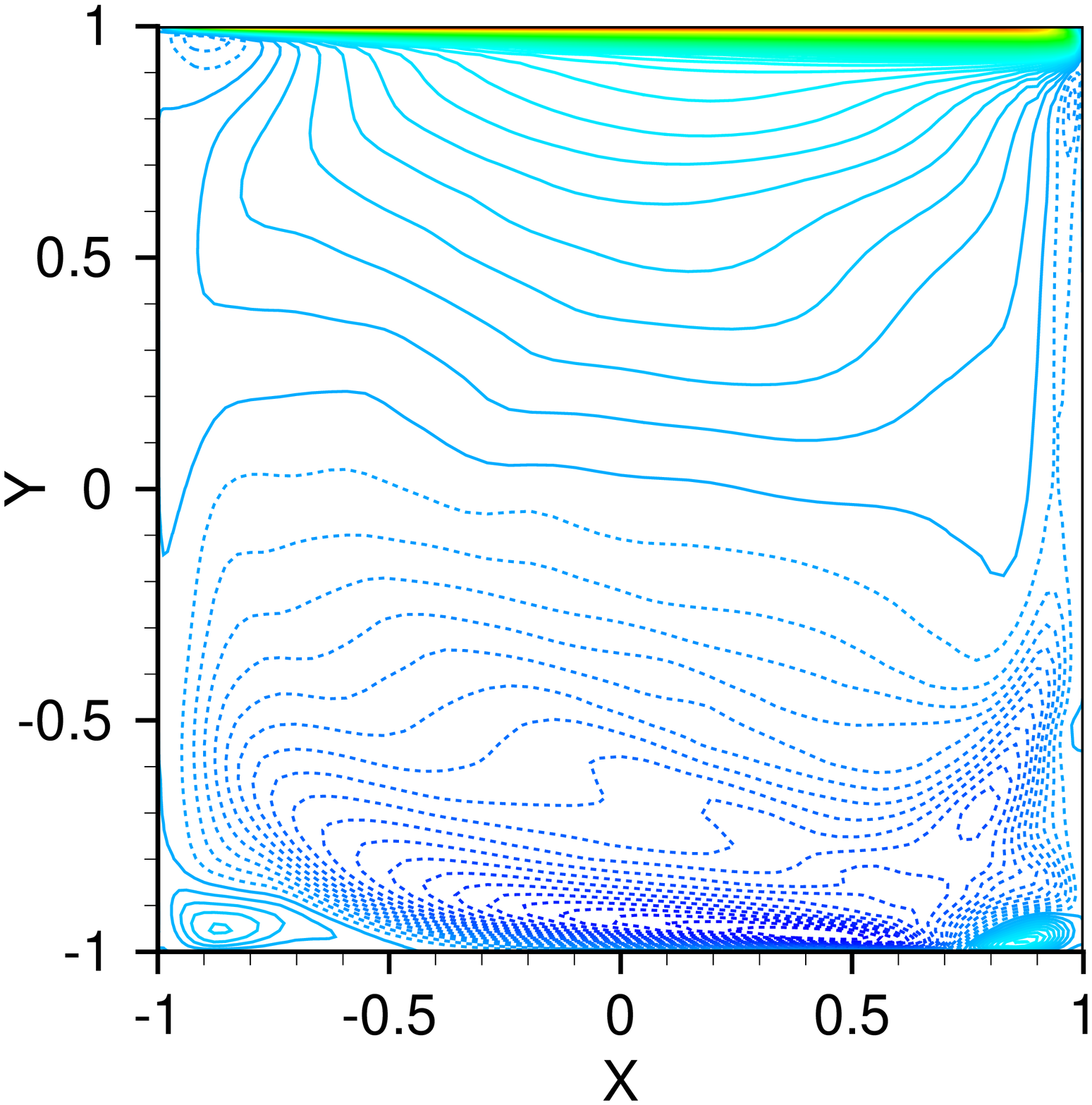}
                \includegraphics[width=0.32\textwidth]{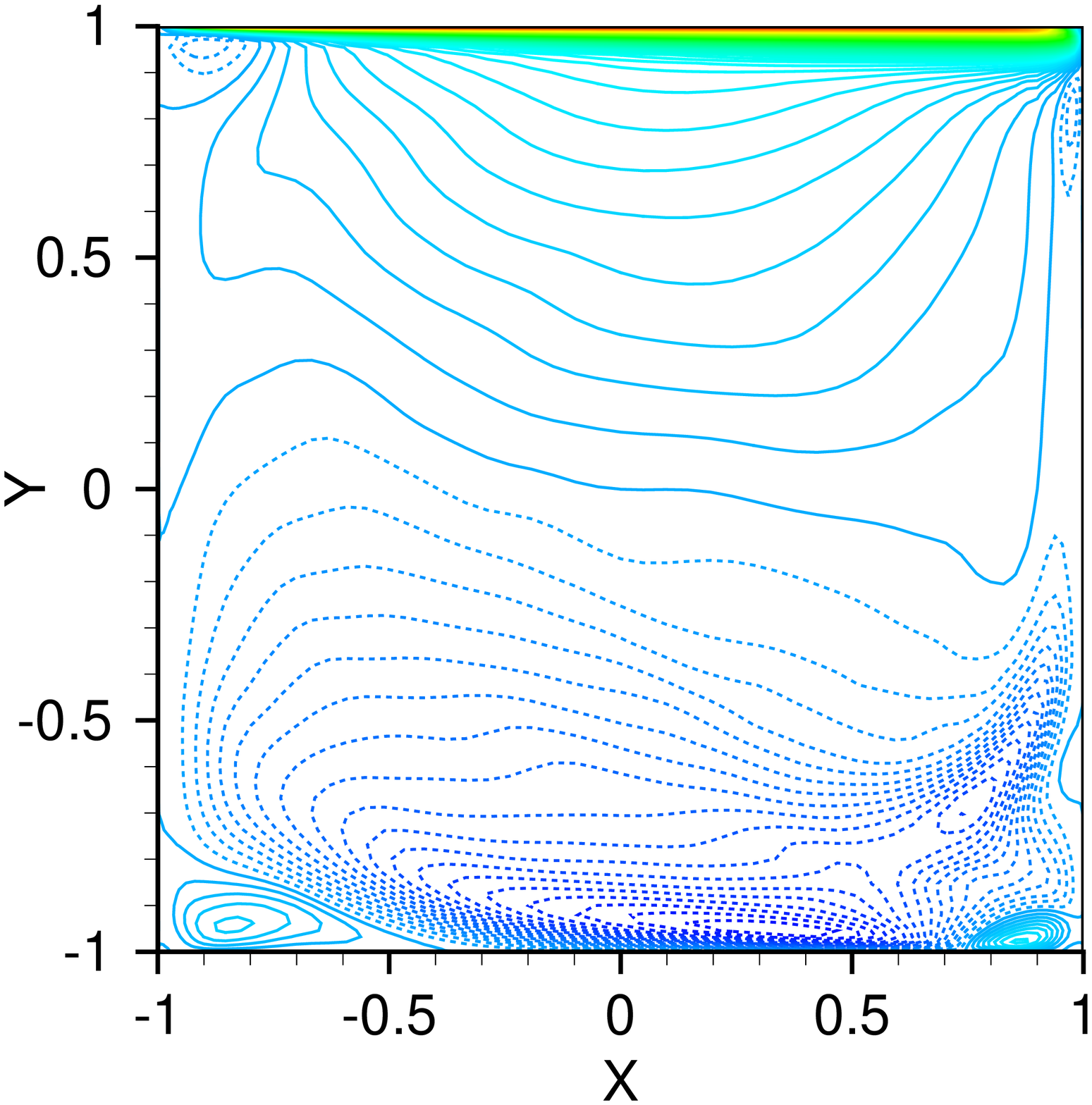}
                \includegraphics[width=0.32\textwidth]{DNS12-um-z0.eps}\\
                \includegraphics[width=0.32\textwidth]{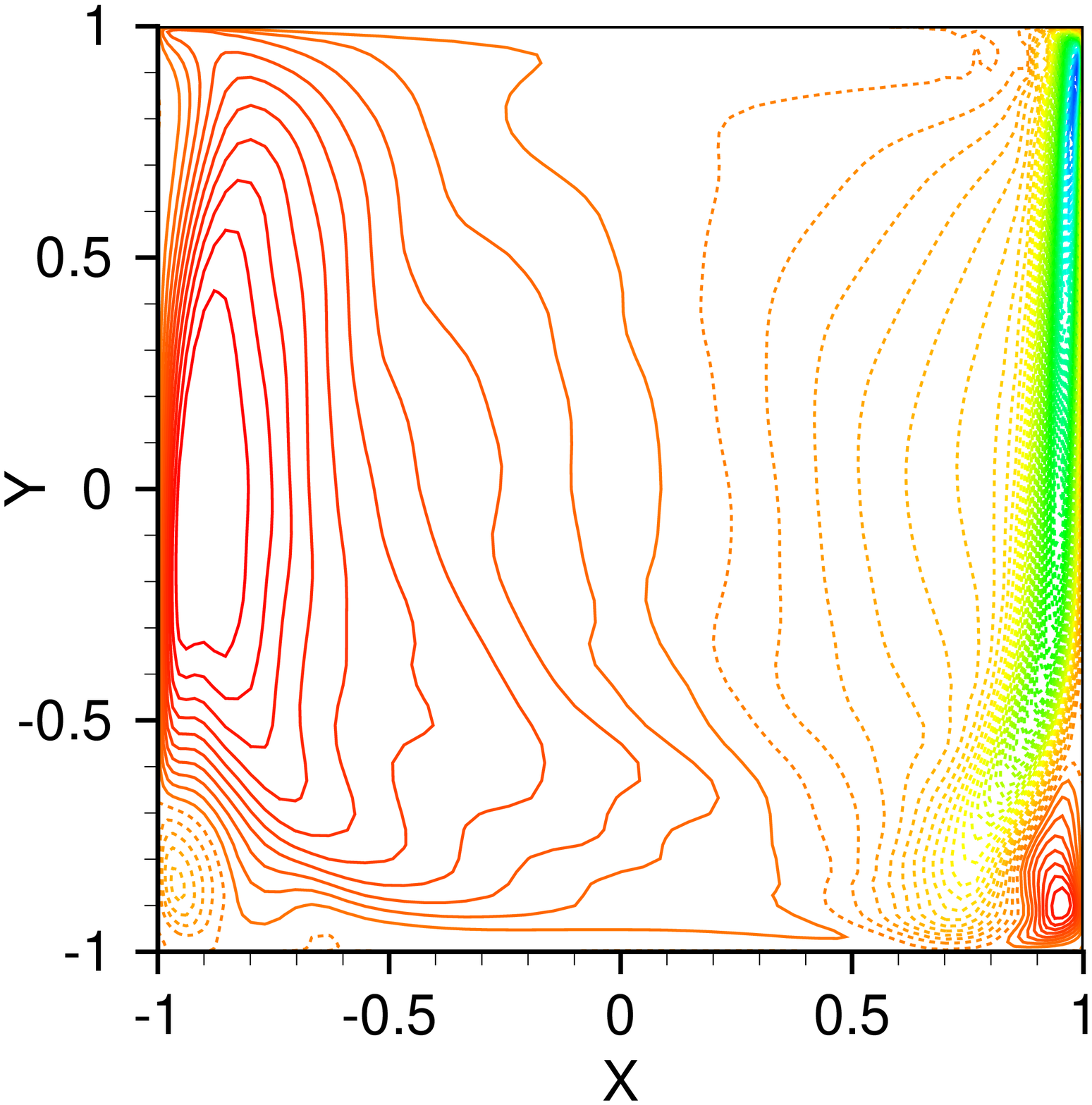}
                \includegraphics[width=0.32\textwidth]{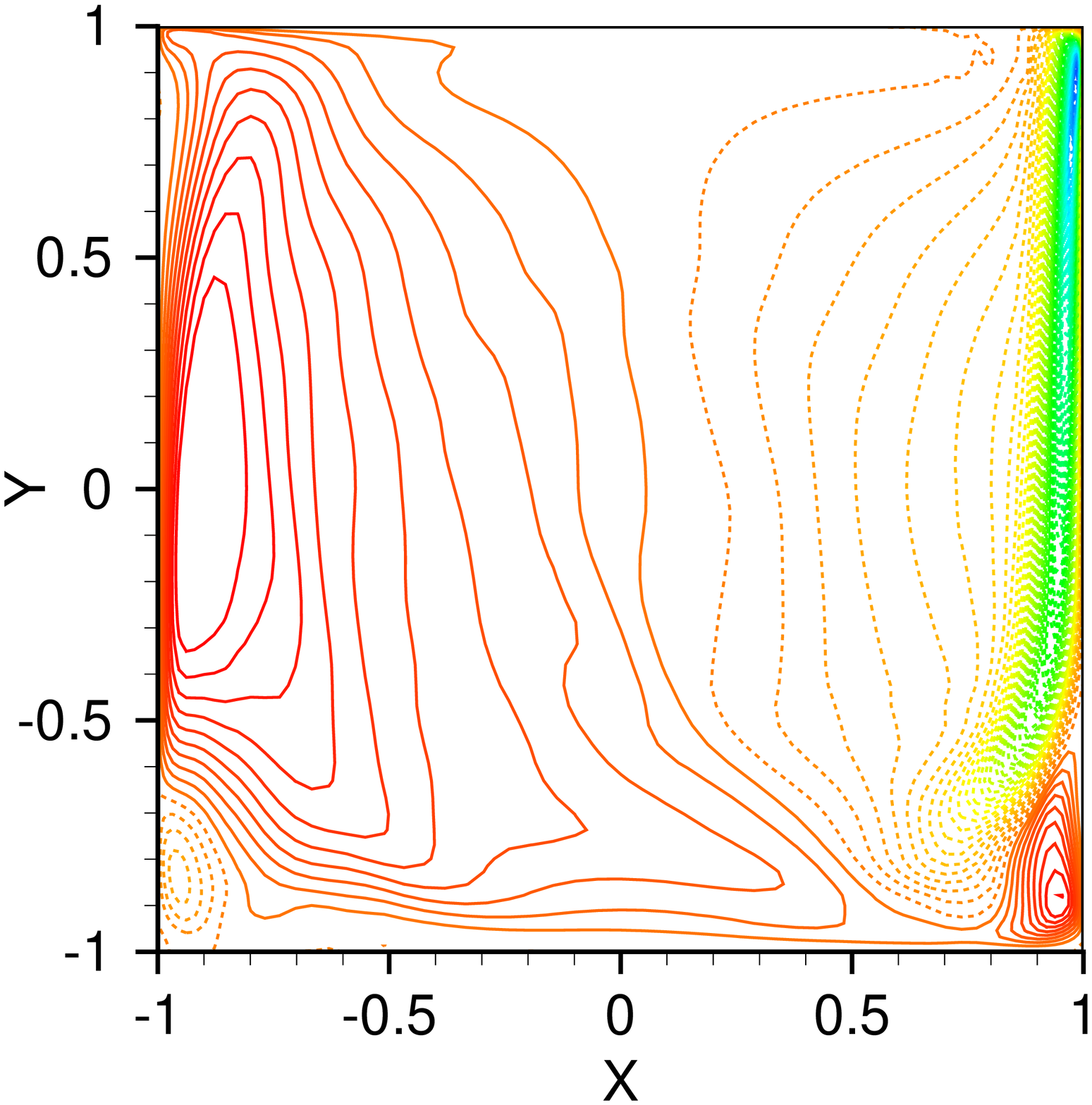}
                \includegraphics[width=0.32\textwidth]{DNS12-vm-z0.eps}
        \caption{In the mid-plane $z/h=0$, DMS (left column), ADM-DMS (central column) and DNS (right column). Top row: contours of $\langle\overline{u}_{x}\rangle$ from $-0.2$ to $1$ by increments of $0.01$. Bottom row: contours of $\langle\overline{u}_{y}\rangle$ from $-0.7$ to $0.1$ by increments of $0.01$. Color scale from blue to red. Dashed contours correspond to negative levels. Levels in $U_0$ units.}
        \label{fig:07-01}
\end{figure}

\begin{figure}[htb!]
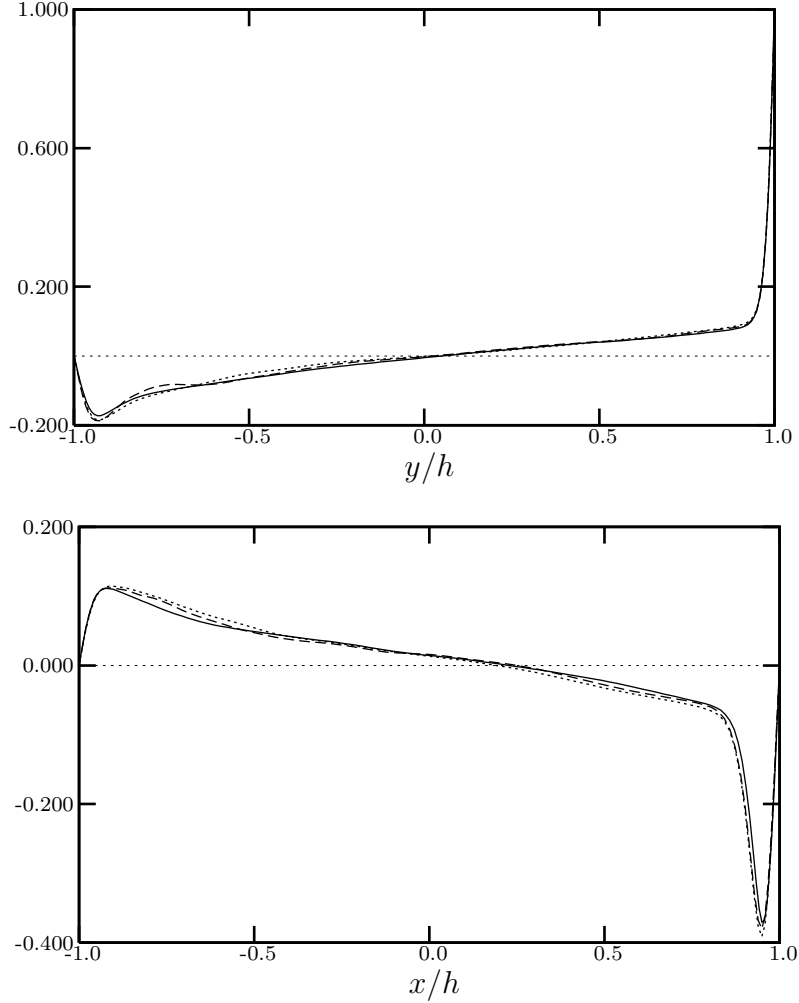

        \centering
        \input{mU-x0-z0.pslatex}
        \vspace{-0.6cm}
        $$ $$
        \input{mV-y0-z0.pslatex}
        \caption{In the mid-plane $z/h=0$, DMS (dashed lines), ADM-DMS (dotted lines) and DNS (solid lines). Top: $\langle\overline{u}_{x}\rangle$ on the horizontal centerline $x/h=0$. Bottom: $\langle\overline{u}_{y}\rangle$ on the vertical centerline $y/h=0$.}
\label{fig:mU-mV}
\end{figure}

\begin{figure}[!ht]
        \centering
                \includegraphics[width=0.32\textwidth]{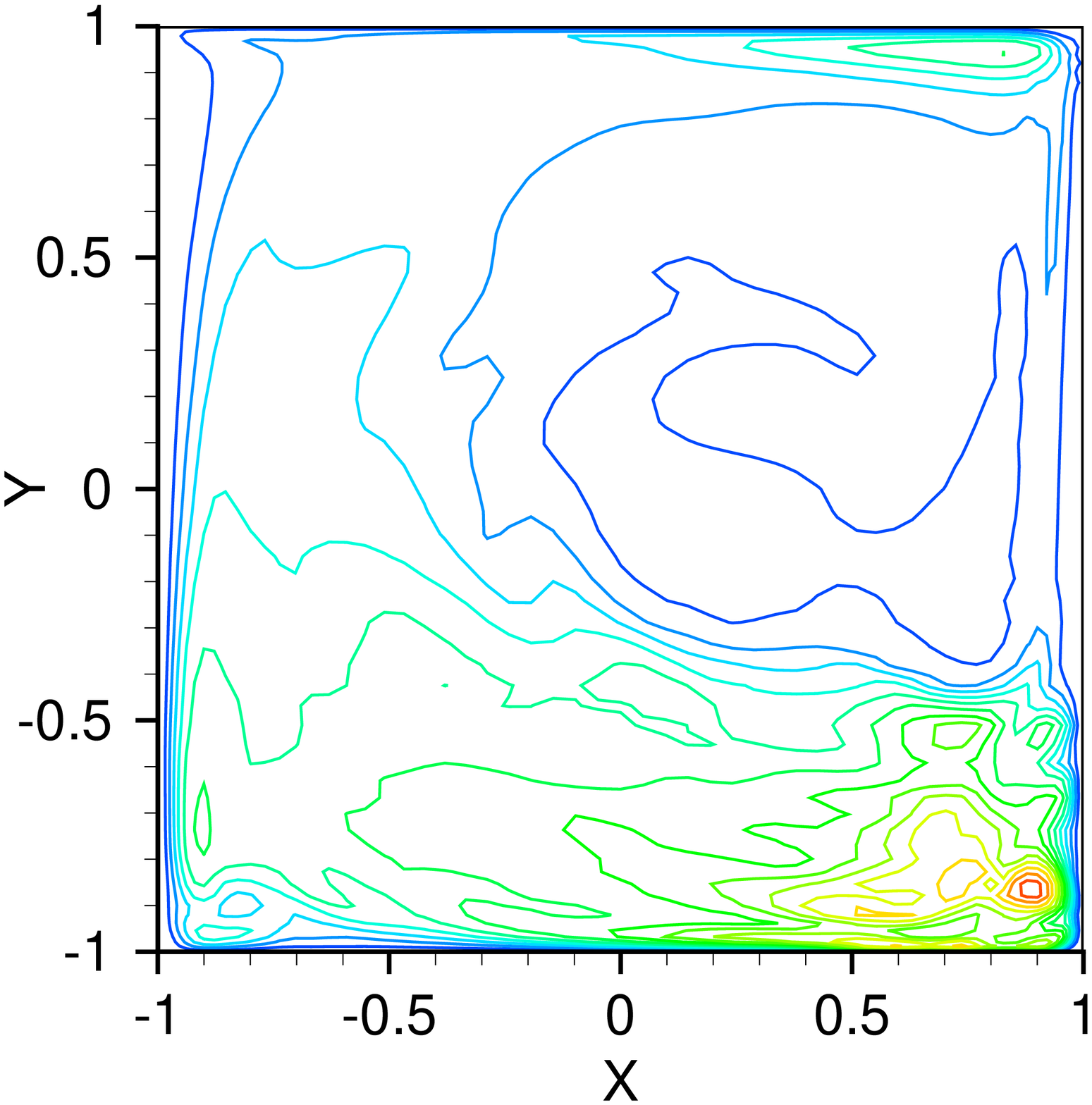}
                \includegraphics[width=0.32\textwidth]{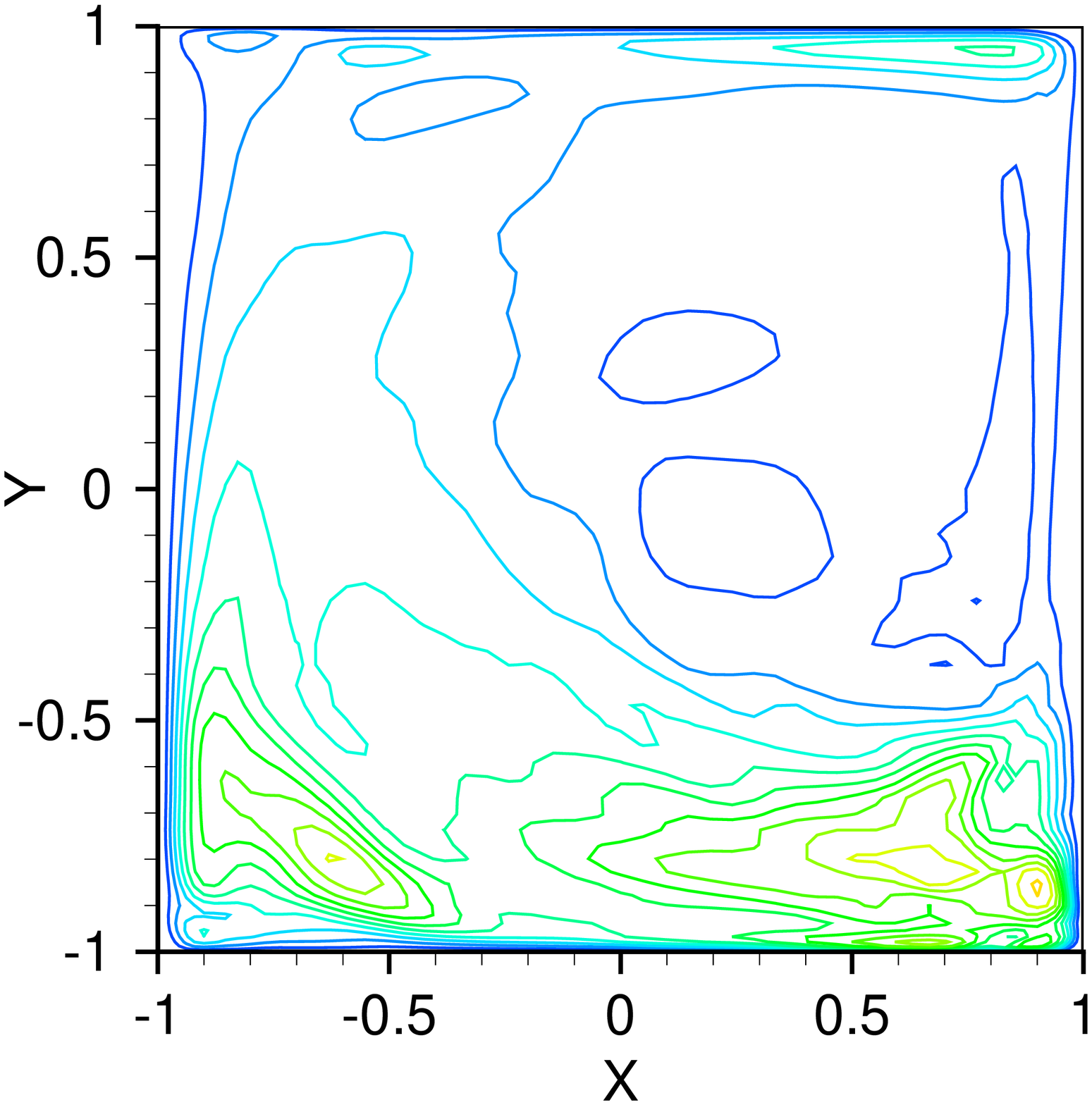}
                \includegraphics[width=0.32\textwidth]{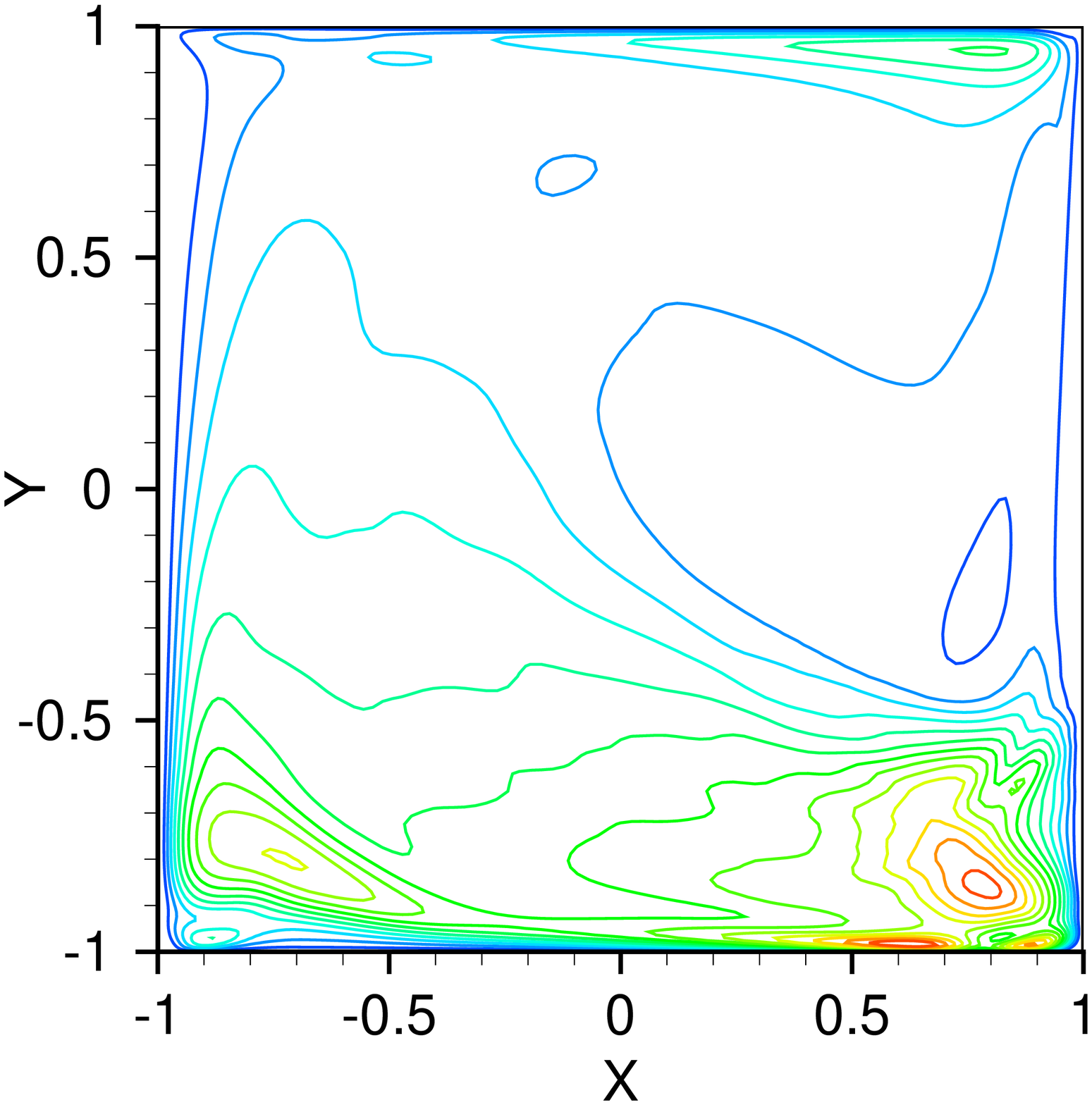}\\
                \includegraphics[width=0.32\textwidth]{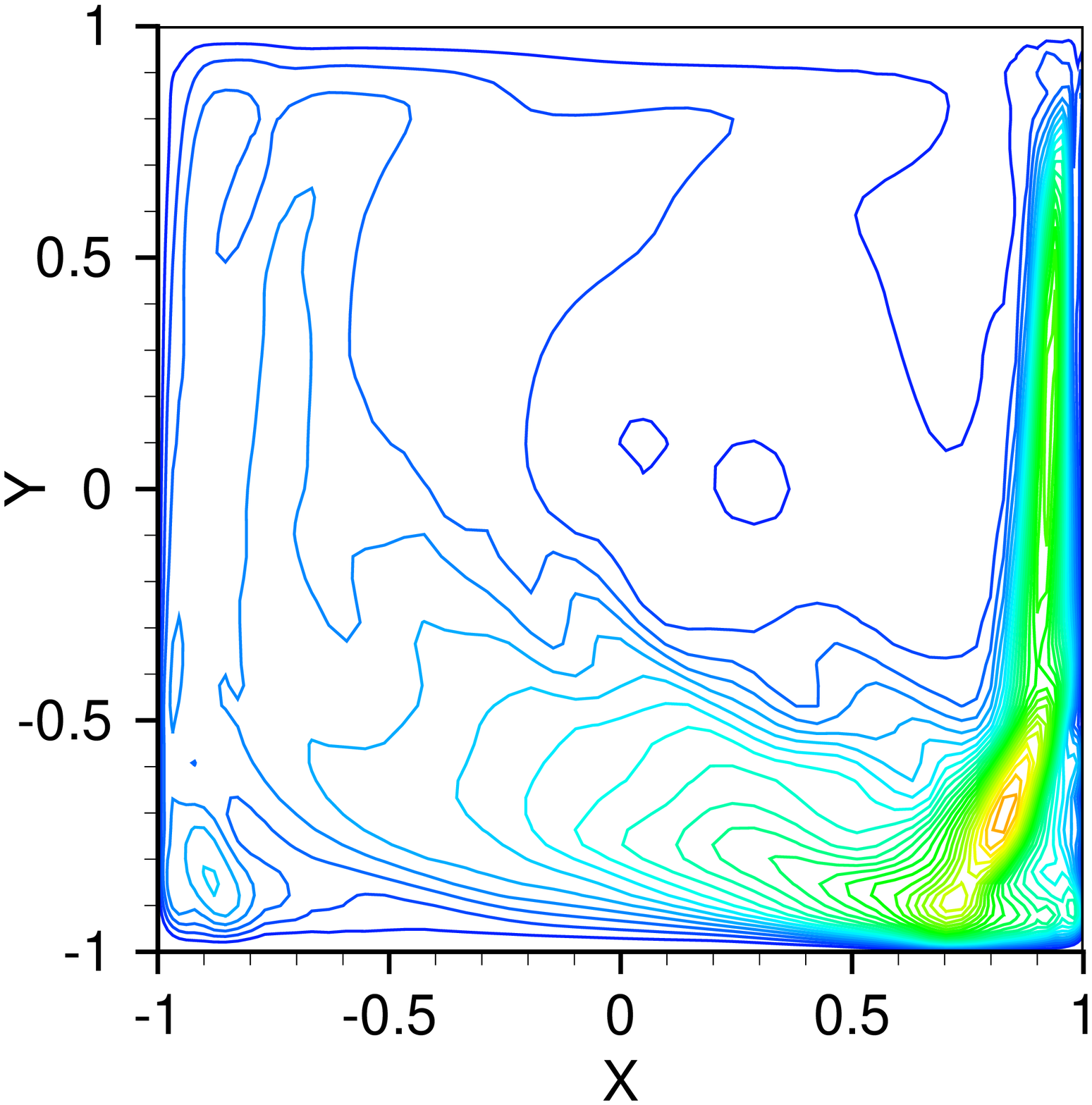}
                \includegraphics[width=0.32\textwidth]{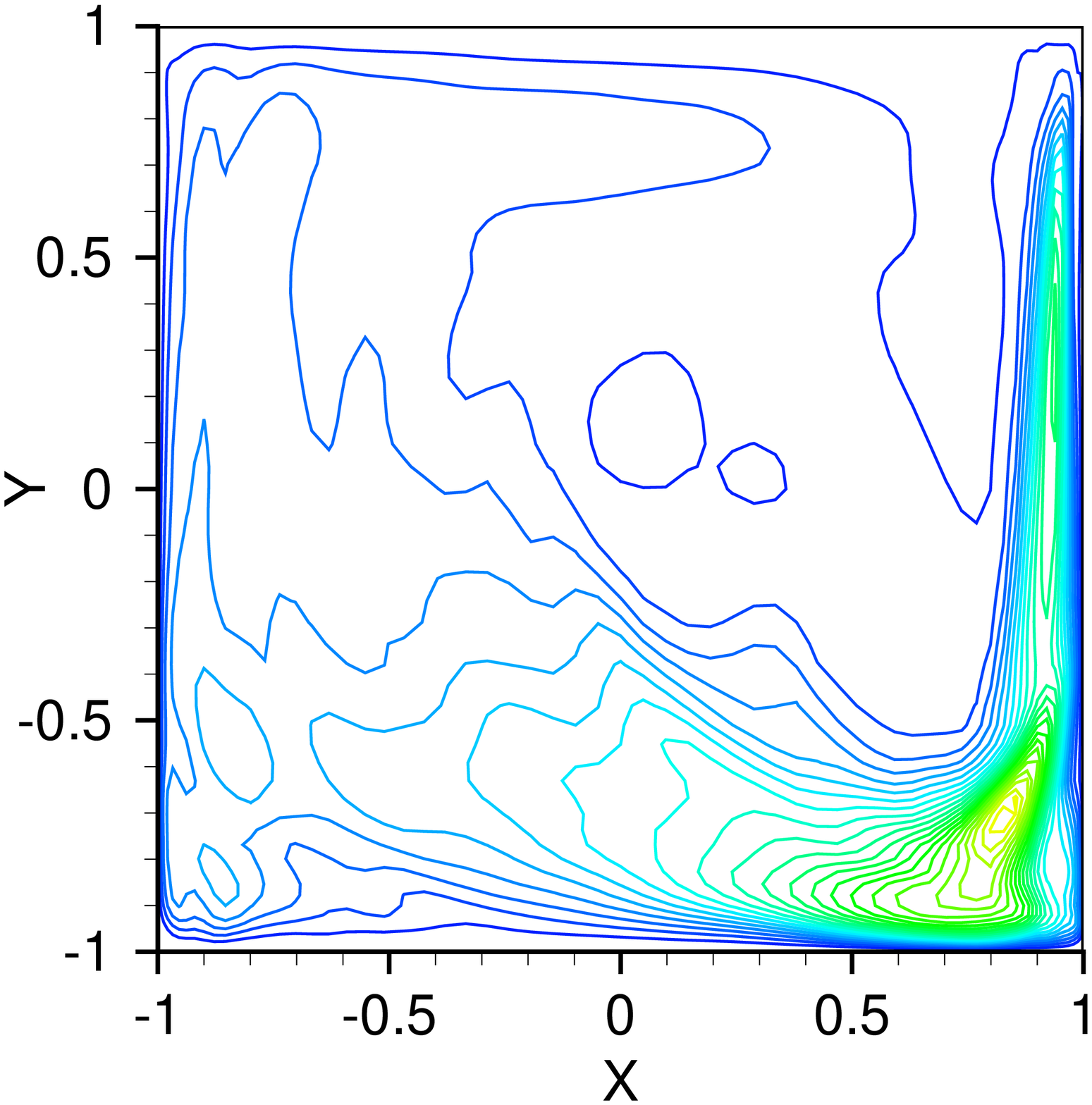}
                \includegraphics[width=0.32\textwidth]{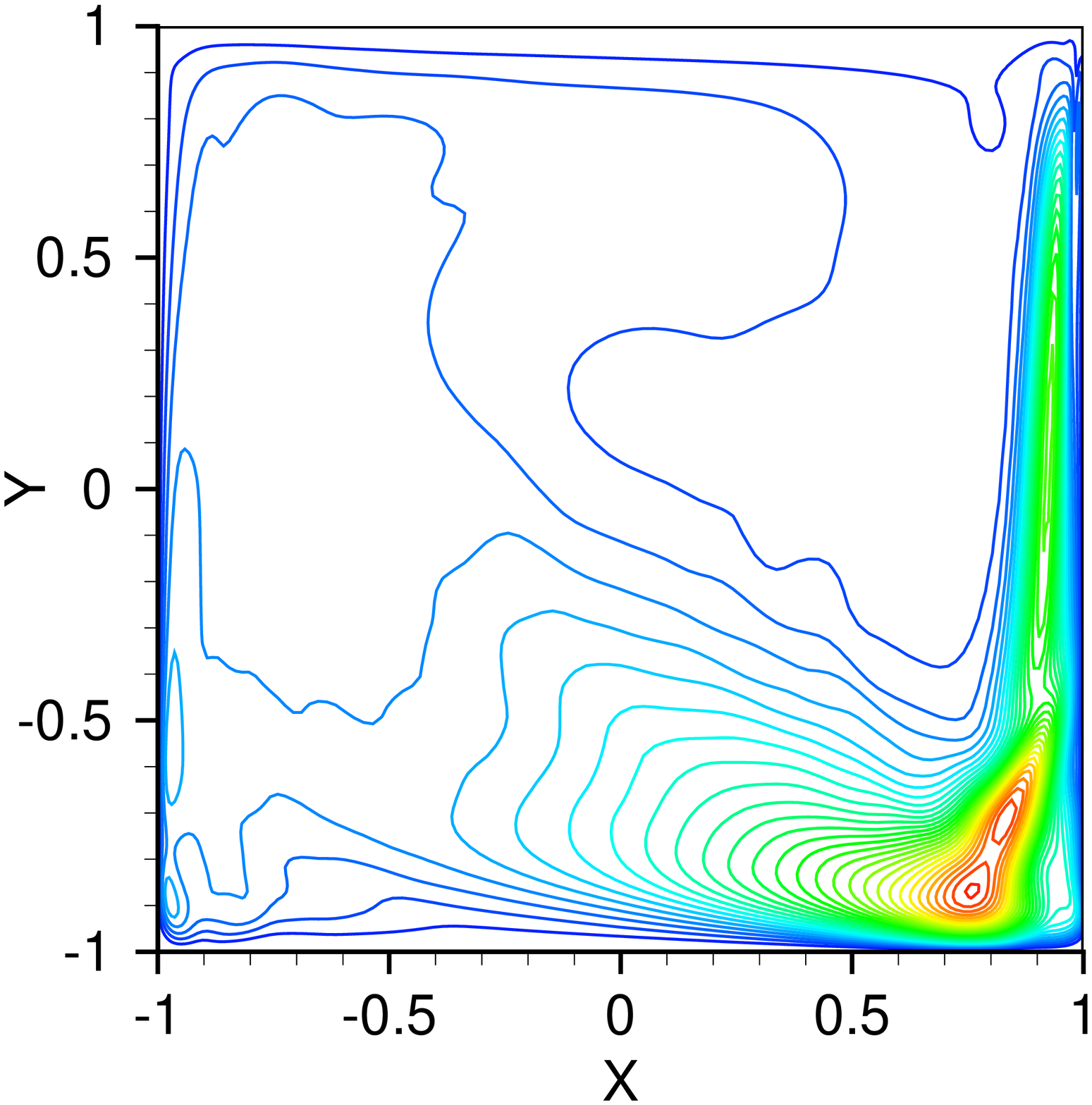}\\
                \includegraphics[width=0.32\textwidth]{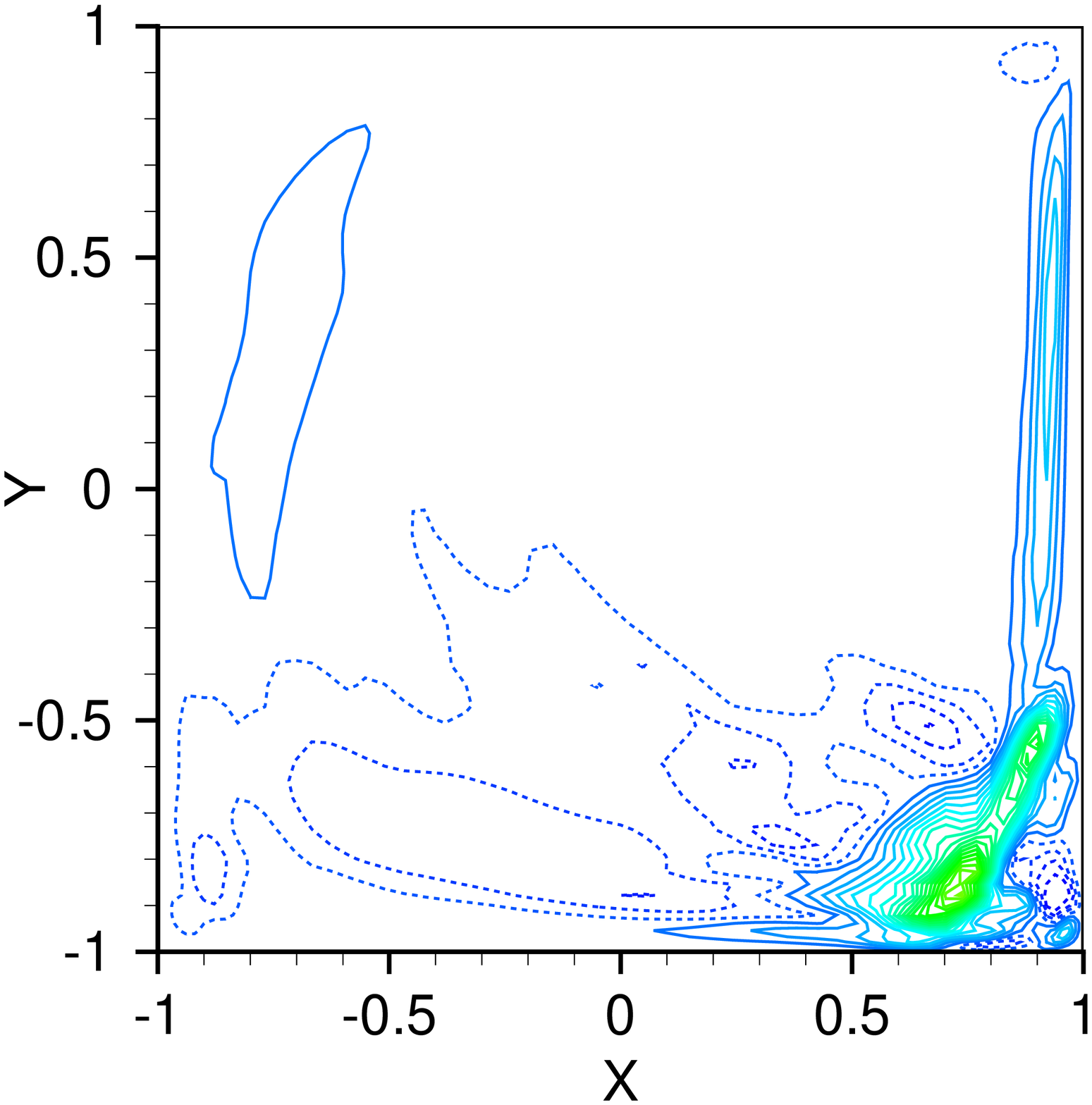}
                \includegraphics[width=0.32\textwidth]{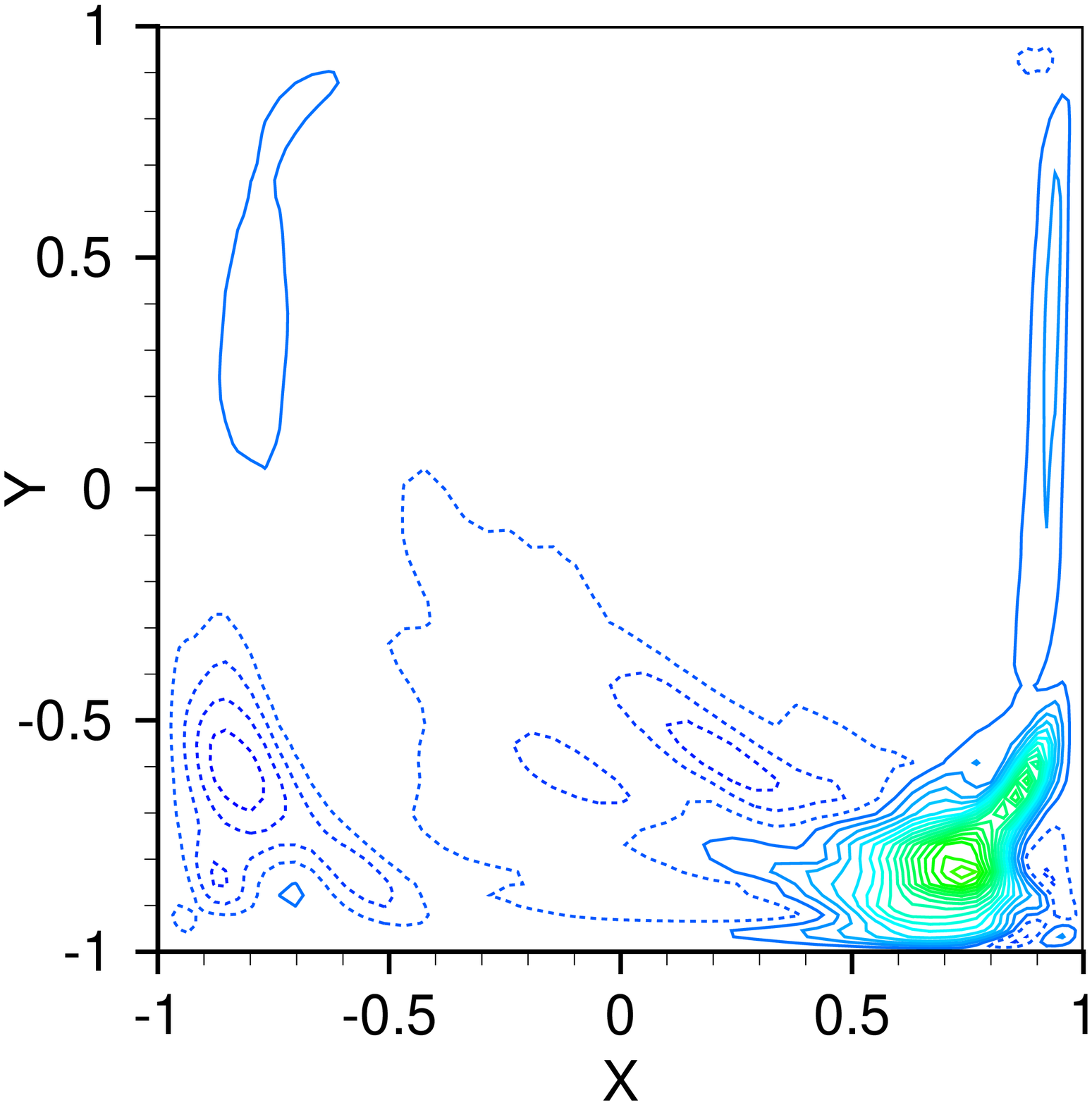}
                \includegraphics[width=0.32\textwidth]{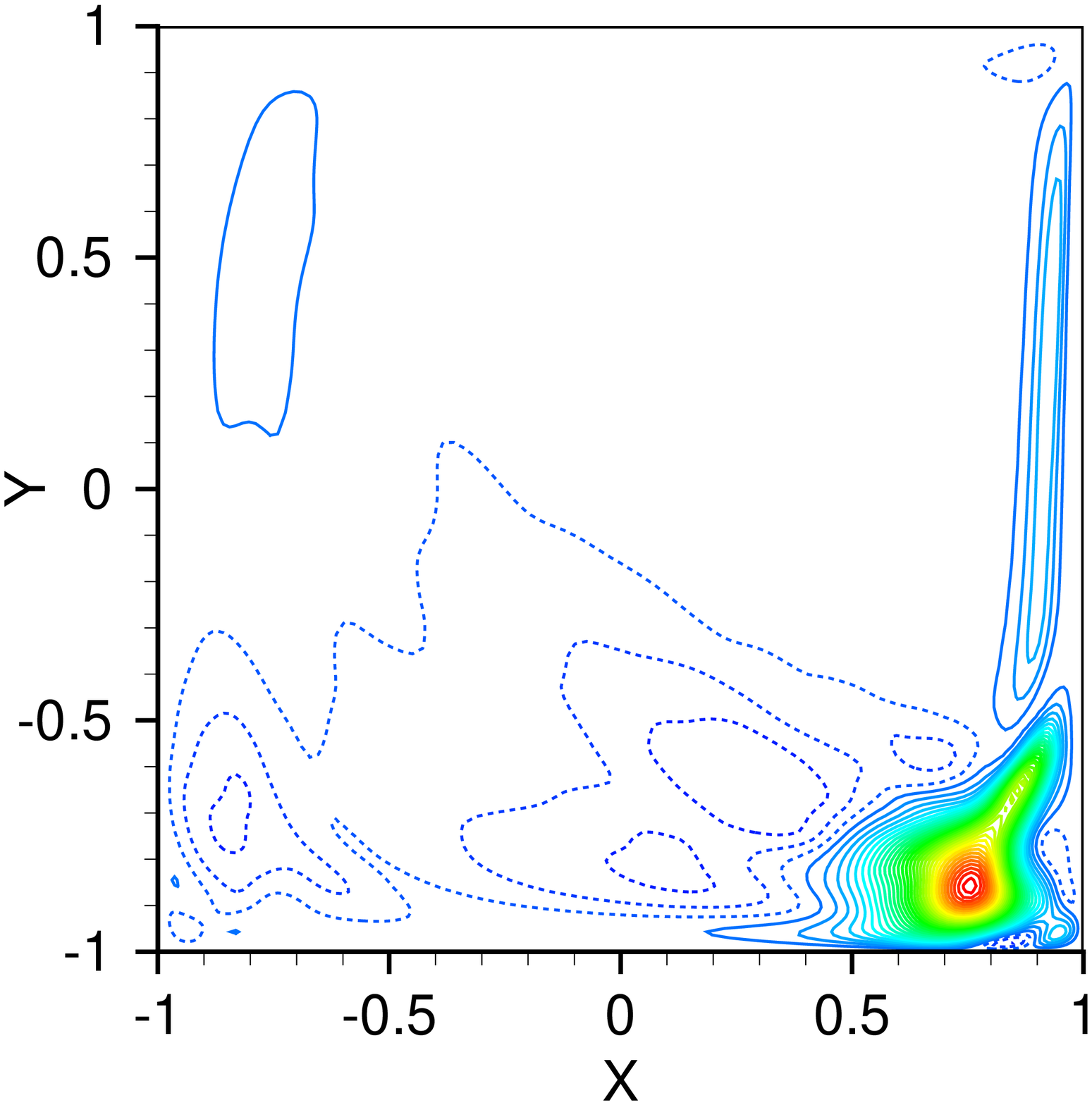}
        \caption{In the mid-plane $z/h=0$, DMS (left column), ADM-DMS (central column) and DNS (right column). Top row: contours of $\langle{\overline{u}_{x}^{\circ}}^2\rangle^{1/2}$ from $0$ to $0.07$ by increments of $0.005$. Central row: contours of $\langle{\overline{u}_{y}^\circ}^2\rangle^{1/2}$ from $0$ to $0.15$ by increments of $0.005$. Bottom row: contours of $\langle\overline{u}_{x}^\circ\overline{u}_{y}^\circ\rangle$ from $-0.0007$ to $0.0065$ by increments of $0.0002$. Color scale from blue to red. Dashed contours correspond to negative levels. Levels in $U_0$ units for $\langle{\overline{u}_{x}^{\circ}}^2\rangle^{1/2}$ and $\langle{\overline{u}_{y}^{\circ}}^2\rangle^{1/2}$ and in $U_0^2$ units for $\langle\overline{u}_{x}^\circ\overline{u}_{y}^\circ\rangle$.}
        \label{fig:07-03}
\end{figure}

The comparisons with the DNS results are performed by plotting identical series of contour levels of the average velocity. Figures \ref{fig:07-01} displays the average values of the velocity field for DMS, ADM-DMS, and the DNS in the mid-plane of the cavity. This figure is complemented by the corresponding one-dimensional plots presented in Fig. \ref{fig:mU-mV} on the horizontal/vertical centerlines in the mid-plane $z/h=0$. A rapid overview of these figures indicates that ADM-DMS provides results very close to the DNS references, which has to be compared with the UDNS results of Figure \ref{fig:07-06}. In addition, it appears that ADM-DMS results are more satisfactory than those from DMS. Secondary corner eddies located above the bottom wall and below the lid next to the upstream wall are correctly resolved in the mean flow. The flow below the lid and near the corner with the downstream wall presents wiggles in the LES contours for  $\langle\overline{u}_{y}\rangle$. More limited effects are noticeable for the equivalent $x$-component field. We assume that these very limited defects find their origin in a local too important under-resolution due to the very high shear rate near the downstream corner right below the lid \cite{bouffanais06:_large}. The previous comparisons of ADM-DMS with the DNS and DMS for first-order moments require to be complemented by plotting identical series of contours of three components of the resolved Reynolds stress tensor. Figure \ref{fig:07-03} showcases the improvement achieved in terms of subgrid modeling by coupling ADM with DMS. Moreover, Fig. \ref{fig:muu}--\ref{fig:muv} provide the associated one-dimensional plots of these quantities in the vertical and horizontal centerlines of the mid-plane of the cavity. Indeed, the variations of $\langle{\overline{u}_{x}^{\circ}}^2\rangle^{1/2}$, $\langle{\overline{u}_{y}^\circ}^2\rangle^{1/2}$ and $\langle{\overline{u}_{x}^{\circ}\overline{u}_{y}^{\circ}}\rangle$ for ADM-DMS reproduce quite accurately the intense-fluctuations zones in the mid-plane $z/h=0$, and more specifically in the vicinity of the downstream corner eddy. DMS appears clearly not as effective as ADM-DMS. The lower intensity of the Reynolds stress components for ADM-DMS as compared to the DNS are induced by the lower sampling of all LES. A longer dynamic range would produce more intense results as reported in \cite{bouffanais06:_large}.

\begin{figure}[htb!]
        \centering
        \input{muu-y0-z0.pslatex}
        \vspace{-0.6cm}
        $$ $$
        \input{muu-x0-z0.pslatex}
        \caption{In the mid-plane $z/h=0$, DMS (dashed lines), ADM-DMS (dotted lines) and DNS (solid lines). $\langle{\overline{u}_{x}^{\circ}}^2\rangle^{1/2}$ on the vertical centerline $y/h=0$ (Top) and on the horizontal centerline $x/h=0$ (Bottom).}
        \label{fig:muu}
\end{figure}

\begin{figure}[htb!]
        \centering
        \input{mvv-y0-z0.pslatex}
        \vspace{-0.6cm}
        $$ $$
        \input{mvv-x0-z0.pslatex}
        \caption{In the mid-plane $z/h=0$, DMS (dashed lines), ADM-DMS (dotted lines) and DNS (solid lines). $\langle{\overline{u}_{y}^{\circ}}^2\rangle^{1/2}$ on the vertical centerline $y/h=0$ (Top) and on the horizontal centerline $x/h=0$ (Bottom).}
        \label{fig:mvv}
\end{figure}

\begin{figure}[htb!]
        \centering
        \input{muv-y0-z0.pslatex}
        \vspace{-0.6cm}
        $$ $$
        \input{muv-x0-z0.pslatex}
        \caption{In the mid-plane $z/h=0$, DMS (dashed lines), ADM-DMS (dotted lines) and DNS (solid lines). $\langle\overline{u}_{x}^\circ\overline{u}_{y}^\circ\rangle$ on the vertical centerline $y/h=0$ (Top) and on the horizontal centerline $x/h=0$ (Bottom).}
        \label{fig:muv}
\end{figure}

\subsection{Reynolds stresses production}
As mentioned by Leriche and Gavrilakis in \cite{leriche00:_direc}, the largest Reynolds stresses production rates in the cavity are to be found in the primary elliptical jets parallel to the downstream wall, near the impact points just above the bottom wall. The budget equations of the resolved second-order moments $\langle\overline{u}_{i}^{\circ}\overline{u}_{j}^{\circ}\rangle$ governing the resolved Reynolds stresses, see \cite{mathieu00:_introd_turbul_flow,pope00:_turbul_flows}, comprise a term named here $\overline{P}_{ij}$, defined by
\begin{equation}
\overline{P}_{ij} = -\langle\overline{u}_{i}^{\circ}\overline{u}_{k}^{\circ}\rangle\frac{\partial\langle\overline{u}_{j}\rangle}{\partial x_{k}}-\langle\overline{u}_{j}^{\circ}\overline{u}_{k}^{\circ}\rangle\frac{\partial\langle\overline{u}_{i}\rangle}{\partial x_{k}}
\end{equation}
and corresponding to the interaction of the mean flow and the Reynolds stress tensor. This quantity can be interpreted as responsible for the production of resolved Reynolds stresses and couples first- and second-order statistical moments.

\begin{figure}[!ht]
        \centering
                \includegraphics[width=0.32\textwidth]{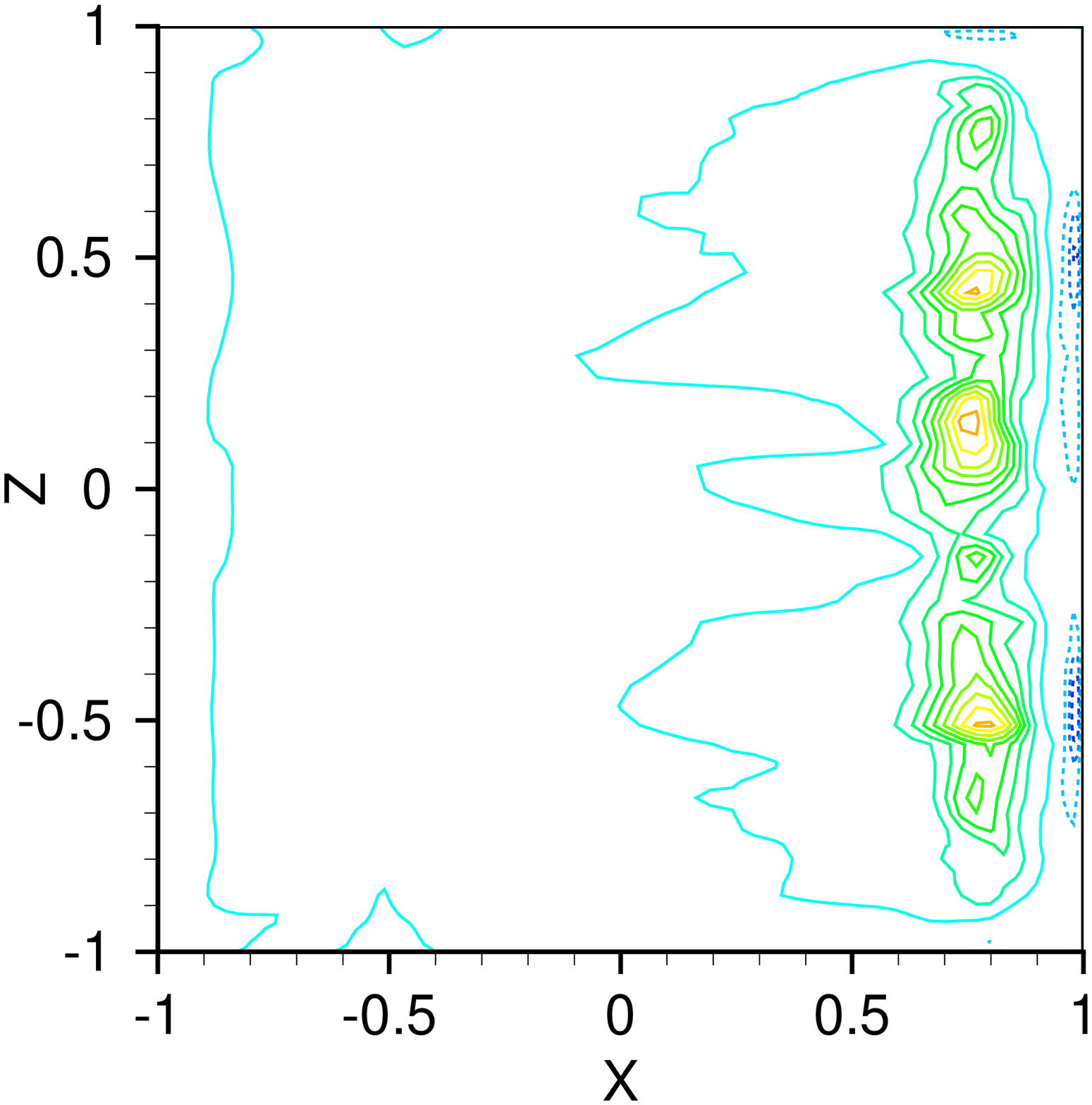}
                \includegraphics[width=0.32\textwidth]{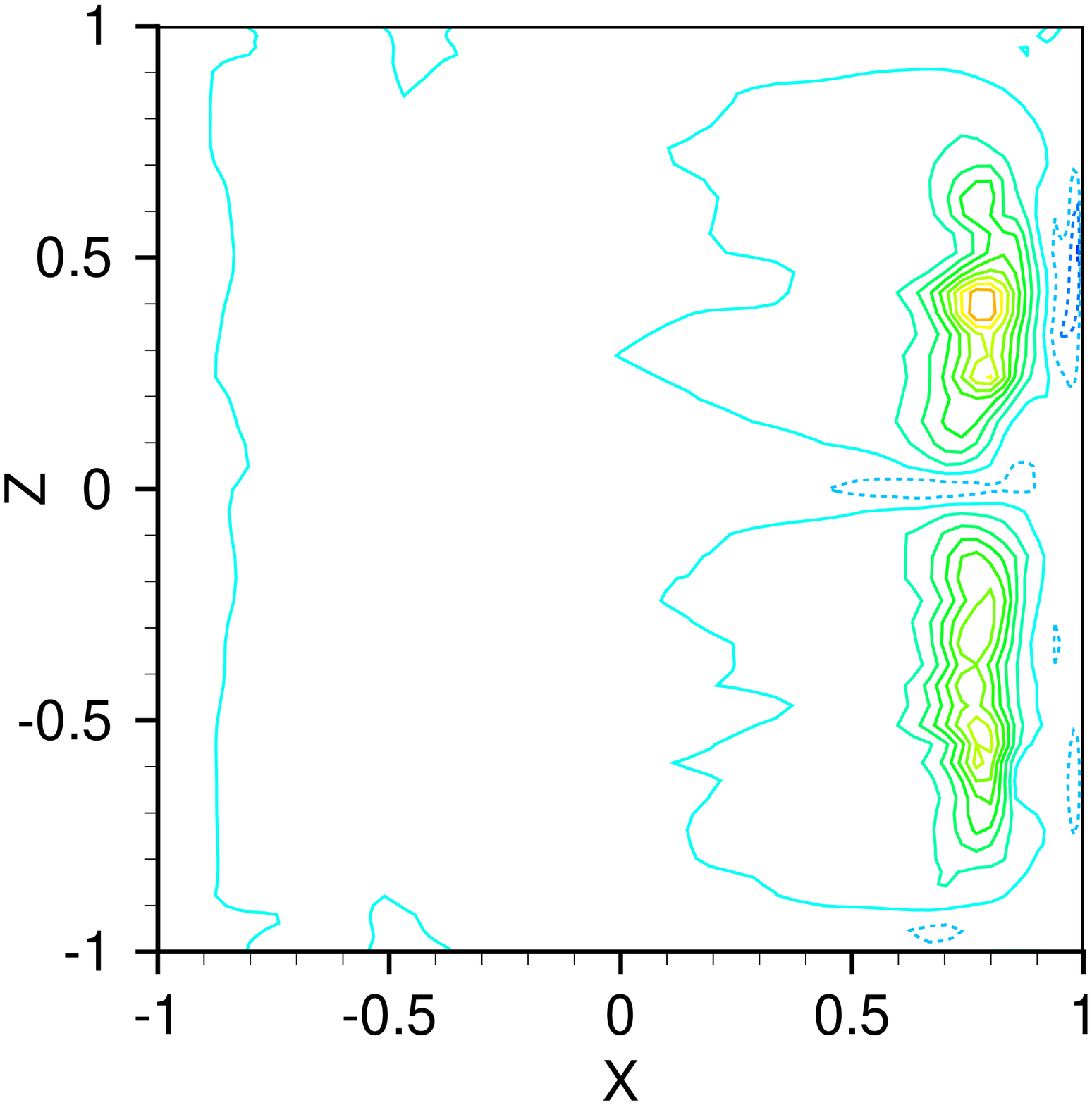}
                \includegraphics[width=0.32\textwidth]{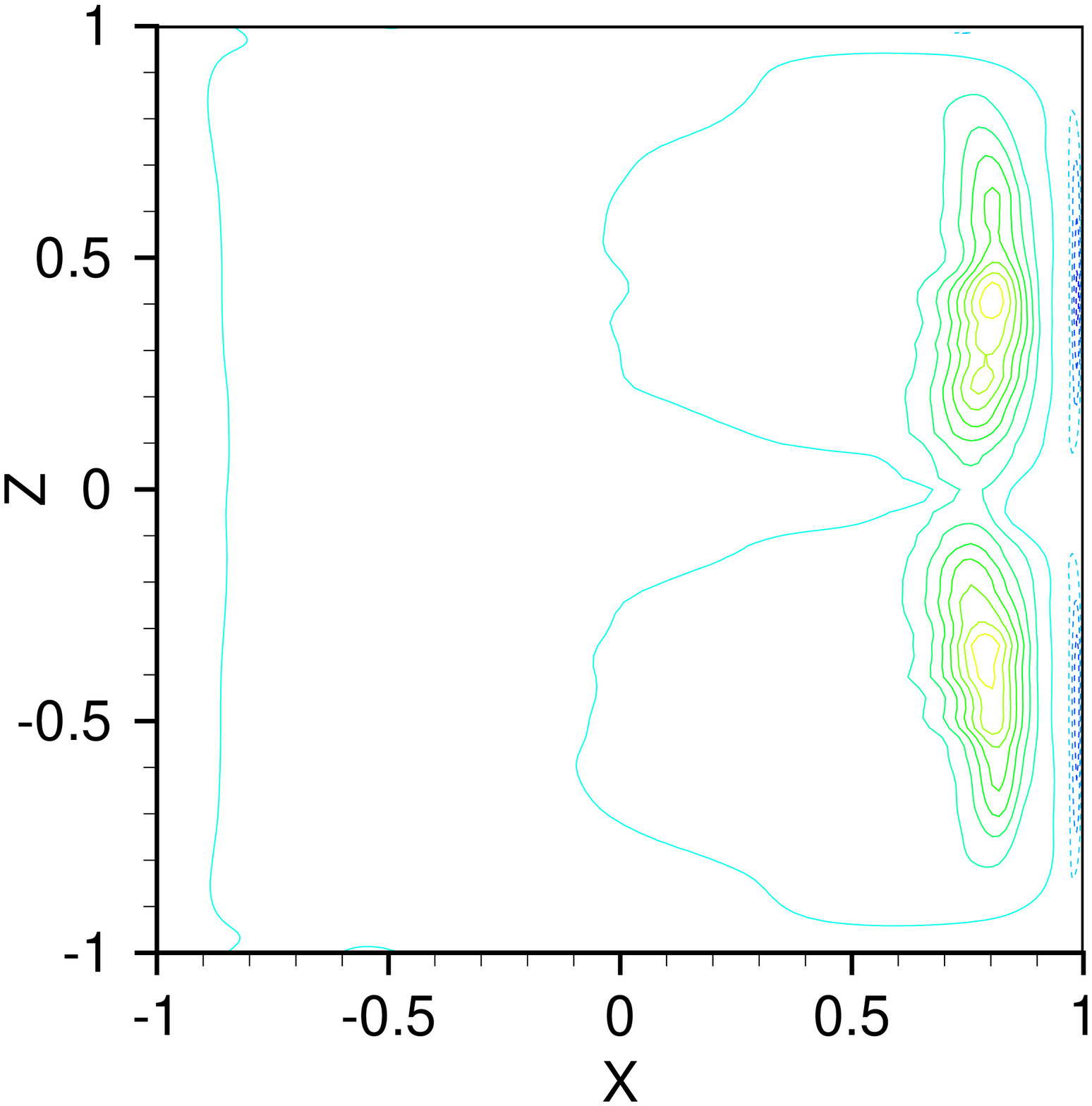}
        \caption{Contours of $\overline{P}_{22}$ from $-0.04$ to $0.11$ by increments of $0.01$. DMS (left), ADM-DMS (center), DNS (right). Plane $y/h=-0.9384$. Color scale from blue to red. Dashed contours correspond to negative levels. Levels in $U_0^3/h$ units.}
        \label{fig:07-09}
\end{figure}

In the specific case of the separated downstream-wall jet, the term $\overline{P}_{22}$ is the largest out of the set of $\overline{P}_{ij}$ terms. After probing in the cavity, the maxima of the field $\overline{P}_{22}$ was found in the plane $y/h=-0.9384$ just at a very short distance above the bottom wall. The contours of the resolved Reynolds stress production term $\overline{P}_{22}$ in this plane are shown in Figure \ref{fig:07-09}. First, it can be noted that these contours are qualitatively very close to the ones obtained by Leriche and Gavrilakis in \cite{leriche00:_direc} and presents secondary structures of negative Reynolds stress production. The distribution of contours allow to clearly visualize the trace of the separated elliptical jets just before their impingement on the bottom wall. This separation is clearer for ADM-DMS than for DMS which shows once again the superiority of the coupled model.

\subsection{Dynamic parameter for ADM-DMS}
In practical LES presented in the sequel, local negative values of the dynamic parameter $\Cd$ are encountered. It was not found necessary to clip them as commonly done---e.g. in \cite{zang93,blackburn03:_spect}---to conveniently get rid of locally destabilizing negative values.

\begin{figure}[htb!]
        \centering
        \includegraphics[width=0.50\textwidth]{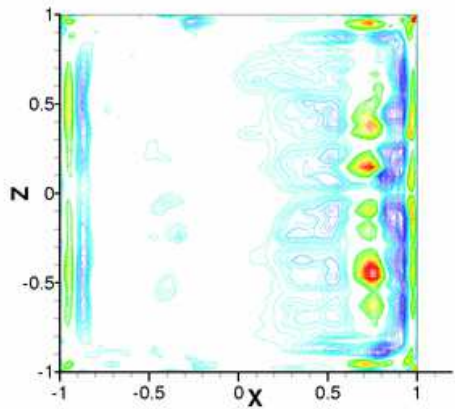}
        \caption{Contours of the average dynamic parameter $\Cd$ from $-0.001$ to $0.001$ for ADM-DMS. Plane $y/h=-0.9384$. Color scale from blue to red. Dashed contours correspond to negative levels.}
        \label{fig:contours-Cd}
\end{figure}

It is worth analyzing the variations of the dynamic parameter $\Cd$ for ADM-DMS in the plane $y/h=-0.9384$ where the maximum of the resolved Reynolds stress production is found. As discussed in Sec. \ref{sec:dynamic}, we expect that, by employing ADM as the base model for the scale-similarity part of the subgrid stress tensor, the magnitude of the dynamic parameter $\Cd$ will be reduced compared with that from the dynamic mixed model and even more reduced compared with that from the dynamic Smagorinsky model \cite{zang93}. This is confirmed by our LES where three orders of magnitude separate the dynamic parameters for DMS and ADM-DMS. The distribution of contours of the average dynamic parameter $\Cd$ in Figure \ref{fig:contours-Cd} appears clearly correlated with the contours of the resolved Reynolds stress production $\overline{P}_{22}$ in the same plane and presented in Figure \ref{fig:07-09}. Indeed, the trace of the separated elliptical jets is discernibly apparent in Figure \ref{fig:contours-Cd}. 

\begin{figure}[htb!]
        \centering
        \input{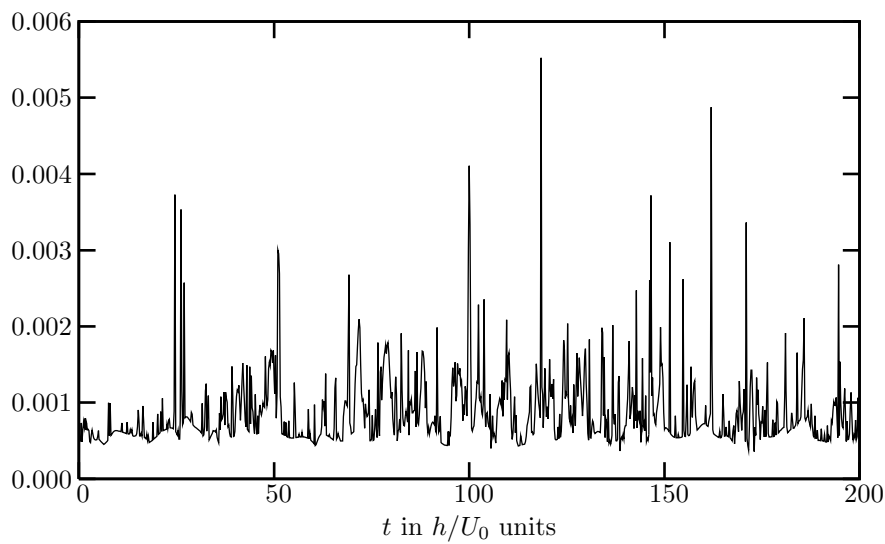}
        \caption{Time history of the local value of the dynamic parameter $\Cd$ for ADM-DMS at the point $\Theta_0$ whose coordinates are $x/h=0.7874$, $y/h=-0.9384$, $z/h=-0.3371$.}
        \label{fig:Cd-max}
\end{figure}

In addition, the maximum of $\overline{P}_{22}$ localized at the point $\Theta_0$ of coordinates $x/h=0.7874$, $y/h=-0.9384$, $z/h=-0.3371$ (see Fig. \ref{fig:07-09}) corresponds to a region of maximal values for the dynamic parameter. The time history of the local value of $\Cd$ at the point $\Theta_0$ is reported in Figure \ref{fig:Cd-max} and present a limited number of high-value peaks. Leriche and Gavrilakis in \cite{leriche00:_direc} and Bouffanais \etal \cite{bouffanais05:_large} identified in this region of the cavity a pair of counter-rotating vortices responsible for the intermittent and intense production of Reynolds stresses. The presence of this coherent vortical structure seems to be detected by the intense values of the dynamic parameter.

\subsection{Subgrid activity}

As a next step, we are mostly interested in identifying the regions where turbulence occurs inside the cavity. For this purpose, we assume that if subgrid scales exist, the flow is locally turbulent and energy is exchanged between subgrid and resolved scales. In other words, the activity of the term modeling subgrid scales is a direct indication of the turbulence occurring in the cavity flow. A measure of subgrid activity is given by the subgrid energy transfer $\overline{\varepsilon}^{\textrm{m}}$ defined by
\begin{equation}\label{eq:06-66}
\overline{\varepsilon}^\textrm{m} = -\tau_{ij}^\textrm{m}\overline{S}_{ij}.
\end{equation}
This latter quantity is only relative in value because the dissipation induced by the fluid viscosity, denoted by $\overline{\varepsilon}_{\nu}$,
\begin{equation}
\overline{\varepsilon}_{\nu} =  2\nu \overline{S}_{ij}\overline{S}_{ij},\qquad \overline{\varepsilon}_{\nu}\geq0,
\end{equation}
is also responsible for an energetic action. It appears therefore legitimate to define and analyze the relative subgrid energy transfer
\begin{equation}
\overline{\varepsilon}_{\mathrm{r}}^{\mathrm{m}} = \frac{\left|\overline{\varepsilon}^{\mathrm{m}}\right|}{\overline{\varepsilon}_{\nu}+\left|\overline{\varepsilon}^{\mathrm{m}}\right|},
\end{equation}
which is referred to as subgrid activity in the sequel. If it is close to zero, the energetic phenomena are mainly induced by the viscous effects showing that the fluid is mainly laminar. Conversely, values close to the unit indicate a strong energetic action of the subgrid model reflecting that turbulence is mainly responsible of energy transfers.

\begin{figure}[!ht]
        \centering
                \includegraphics[width=0.32\textwidth,angle=270]{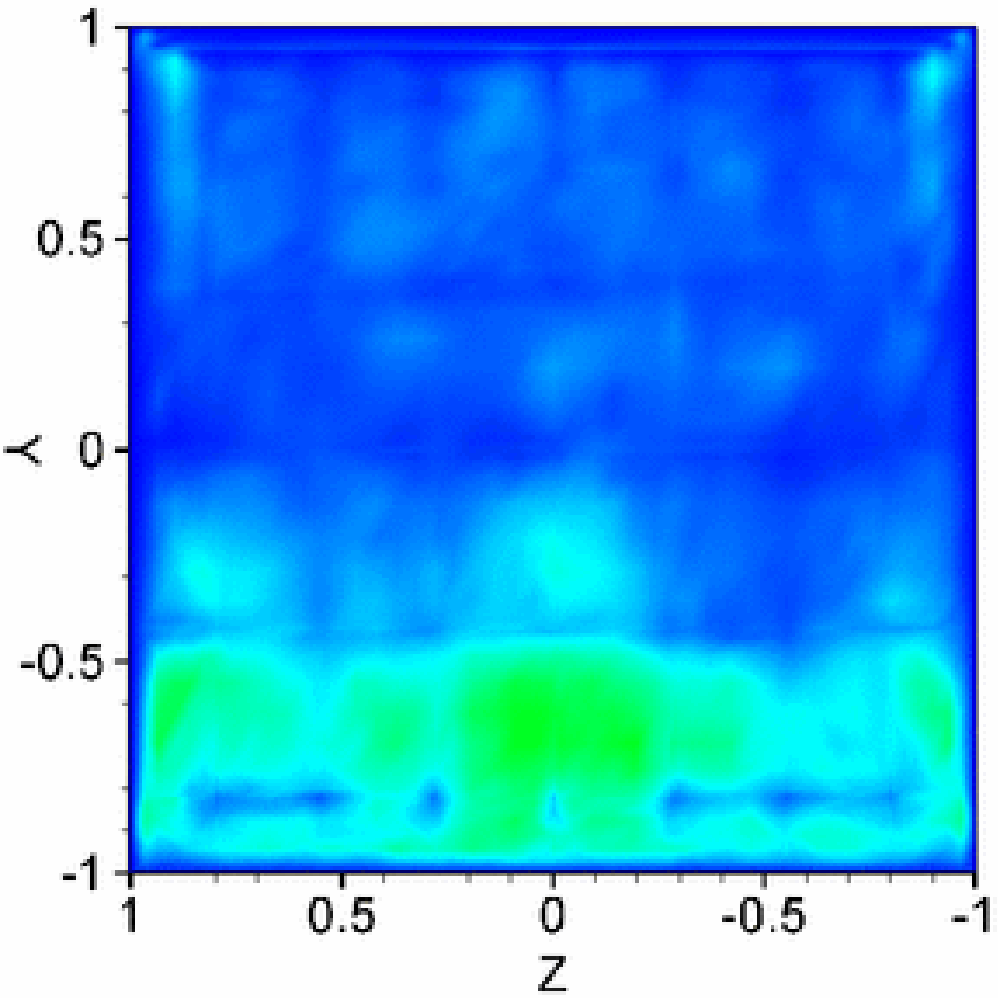}
                \includegraphics[width=0.32\textwidth,angle=270]{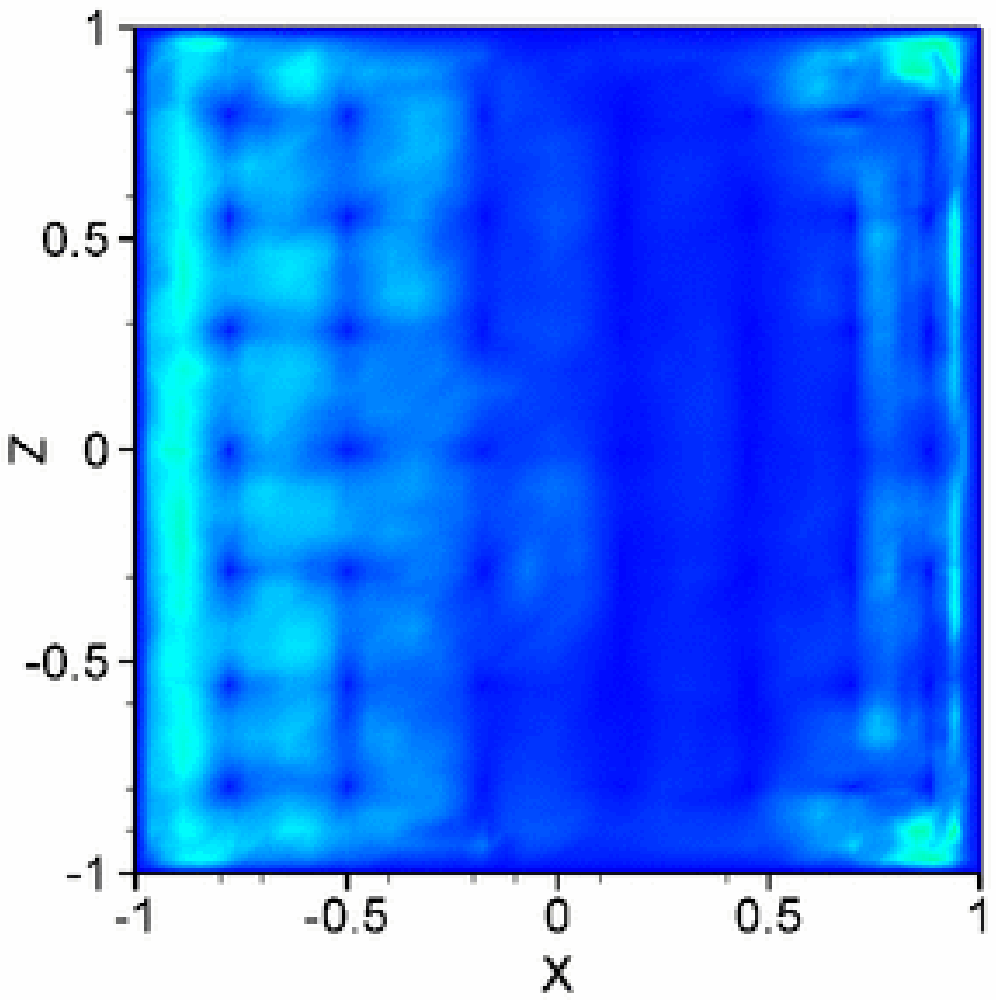}
                \includegraphics[width=0.32\textwidth,angle=270]{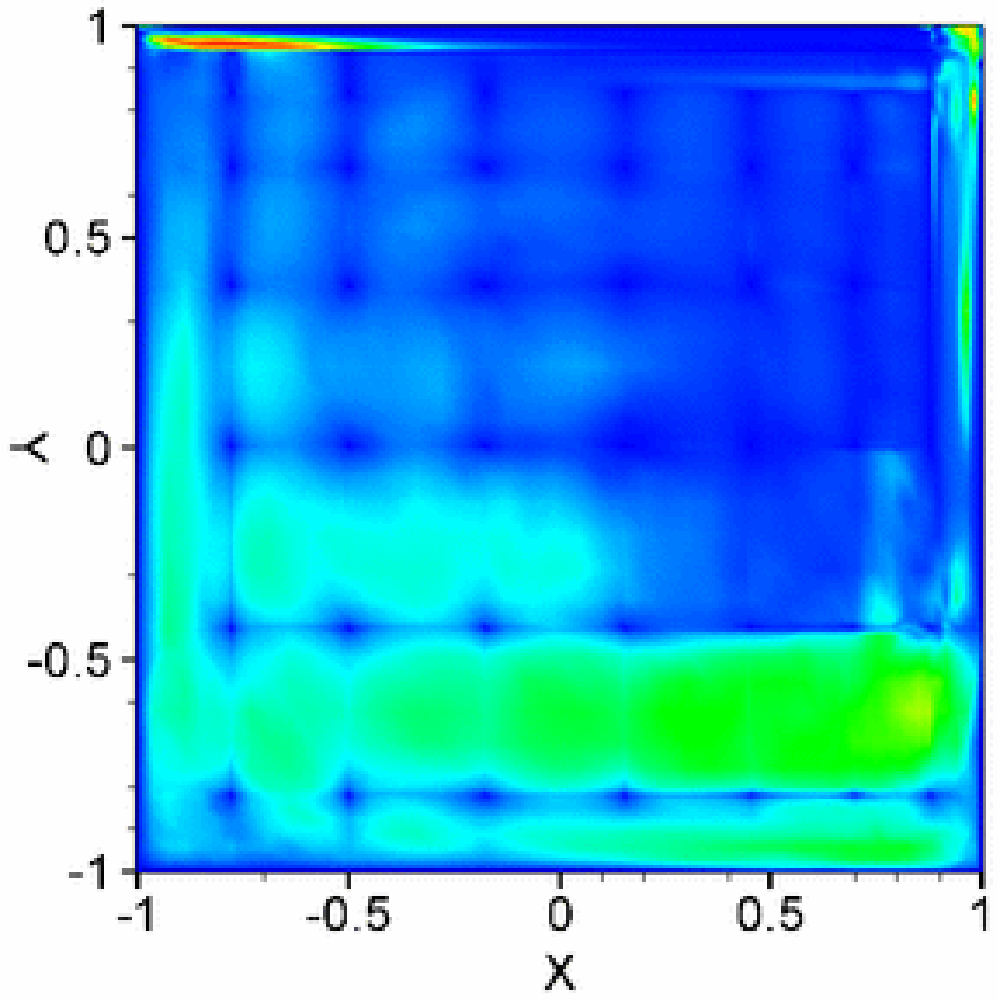}
        \caption{Map of the average relative subgrid energy transfer $\langle\varepsilon_{\mathrm{r}}\rangle$ from $0$ (blue) to $1$ (red) for ADM-DMS. Upstream wall $x/h=0$ (left), bottom wall $y/h=0$ (center) and mid-plane $z/h=0$ (right).}
        \label{fig:07-08}
\end{figure}

As one can see in Fig. \ref{fig:07-08}, turbulence essentially occurs in the vicinity of the cavity walls but, as expected, very close to the walls the energetic action is essentially due to viscous effects. High values of subgrid activity are also identified at the bottom of the cavity and near upstream and downstream walls. One can also notice that subgrid activity is clearly reduced at the edges of the elements. This is a direct consequence of the nature of the filter which is not active at the element-boundaries. This issue cannot be avoided in this framework since $C^{0}$-continuity of the variables across elements is essential for numerical stability and physical consistency reasons.

\subsection{Subgrid kinetic energy}
\begin{figure}[htb!]
        \centering
        \includegraphics[width=0.50\textwidth]{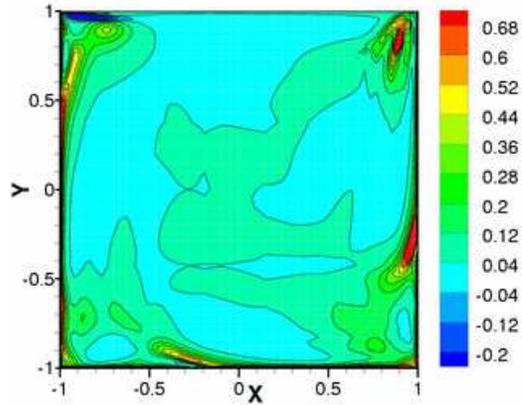}
        \caption{Map of the ratio between the resolved and the subgrid kinetic energy in the plane $z/h=0.9384$.}
        \label{fig:kappa}
\end{figure}

In order to complement the previous study of the subgrid activity and give further details about the importance of the subgrid terms in the ADM-DMS simulation, energetic quantities related to the subgrid scales are analyzed. For this purpose, we consider the filtered subgrid kinetic energy $\overline{q}^{\prime}$ which is expressed---see \cite{sagaut03:_large,pope00:_turbul_flows}---as the difference between the total filtered kinetic energy and the kinetic energy of the resolved field $\overline{q}=\overline{u}_i\overline{u}_i/2$,
\begin{equation}
  \overline{q}^{\prime}   = \frac{1}{2}(\overline{u_{i}u}_{i}-\overline{u}_i\overline{u}_i) = \frac{1}{2}\tau_{ii} \simeq \frac{1}{2}\tau_{ii}^{\mathrm{m}},
\end{equation}
where $\btau^{\textrm{m}}$ is the modeled subgrid tensor defined in Eq. \eqref{eq:04-16c}. In order to provide the reader with deeper insight into the relative importance of the subgrid terms, we introduce the relative subgrid kinetic energy $\kappa$ as the ratio between the subgrid kinetic energy and the kinetic energy of the resolved field
\begin{equation}
  \kappa = \frac{\overline{q}^{\prime}}{\overline{q}}. 
\end{equation}
As one can notice on Fig. \ref{fig:kappa}, the average values of $\kappa$ reported in the plane $z/h=0.9384$, have negative values mainly located at the top-left corner of the cavity. This shows that the subgrid model predicts backscattering, that is the energy transfer from subgrid to resolved scales. This region of inverse energy transfer corresponds to the region of intense subgrid activity as reported in Fig. \ref{fig:07-08} in the mid-plane $z/h=0$. 

The importance of the subgrid terms already observed for the local energy fluxes and analyzed through the subgrid activity, is further confirmed by the presence of regions of intense $\langle \kappa \rangle$. More precisely, four regions with high values of $\langle\kappa\rangle$ and corresponding to the zones where the wall-jets detach from their respective walls \cite{leriche00:_direc} are easily identified in Fig. \ref{fig:07-08}.

\begin{figure}[htb!]
        \centering
        \includegraphics[width=0.50\textwidth]{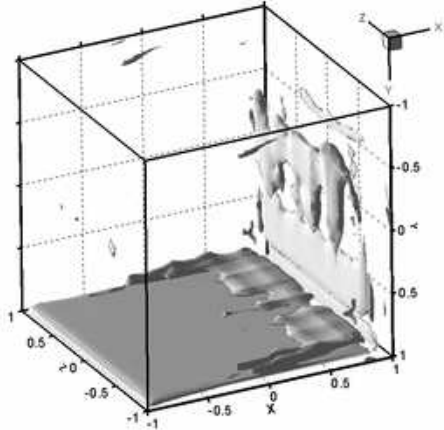}
        \caption{Iso-surface of vanishing subgrid kinetic energy in the cavity.}
        \label{fig:back}
\end{figure}

In order to highlight the zones where backscattering occurs, the iso-surface $\langle \overline{q}^{\prime}\rangle = 0$ is plotted in Fig. \ref{fig:back} thereby defining the boundary between the backscattering and the forward-energy transfer regions. This figure shows that this phenomenon mainly occurs below the lid and in the down flowing jet next to the downstream wall. It also demonstrates the need for a complex subgrid model with such anisotropic flows containing various flow conditions and no direction of homogeneity.

\section{Conclusions}
LES of Newtonian incompressible fluid flows with ADM based on the van Cittert method using Legendre-SEM\ have been performed. A coupling with a dynamic mixed scale model was introduced. The coupling of the lid-driven cubical cavity flow problem at Reynolds number of 12'000 with the SEM\ having very low numerical dissipation and dispersion appears to be a well suited framework to analyze the accuracy of the proposed subgrid model.

The filtering operation is performed in a spectral modal space, generated by a hierarchical basis using the Legendre polynomials, through the application of a specifically designed transfer function. This transfer function is constructed in order to ensure continuity across elements, conservation of the constants, invertibility of the filter and to perform low-pass filtering. From the computation viewpoint, the filtering technique presented in this article, is the essential link between the SEM and ADM-based subgrid models.

The validation of the deconvolution procedure performed using a DNS velocity sample, shows that the van Cittert method is convergent. Accounting for the reduced sampling and integration time, the LES performed with  ADM-DMS show good agreement with the reference results. More precisely, first- and second-order statistics are in good agreement when compared to their DNS counterparts. Results for the Reynolds stresses production, coupling first- and second-order statistical moments, are also well predicted using this new model even with such reduced sampling. The analysis of the results obtained with DMS allows us to clearly identify the improvement induced by coupling ADM with DMS. Subgrid activity has been analyzed showing a qualitative correlation with the localization of small-scale structures in the cavity depicted in \cite{bouffanais06:_large}. The importance of the subgrid kinetic energy as compared to the kinetic energy of the resolved field highlights the essential need for an appropriate subgrid modeling. Furthermore, regions of backscatter are identified by ADM-DMS.

All the presented results emphasize the efficiency of ADM-DMS when dealing with laminar, transitional and turbulent flow conditions such as those occurring in the lid-driven cubical cavity flow at $\textrm{Re}=12'000$. 

%Acknowledgments
\ack
The authors would like to thank Dr. Roland von Kaenel from CFS Engineering, Lausanne, for insightful discussions.

This research is being partially funded by a Swiss National Science Foundation Grant (No. 200020--101707), whose support is gratefully acknowledged.

The DNS data were obtained on supercomputing facilities at the Swiss National Supercomputing Center CSCS and the LES data on Pleiades and Pleiades2 clusters at EPFL--ISE.
\newpage
%Bibliography
%\bibliography{/data/rbouffan/Research_Project/Thesis/biblio_main_bibtex/biblio.bib}

\end{document}